\documentclass[journal]{IEEEtran}

\usepackage{color,soul} %to highlight a text
\usepackage{todonotes}

\usepackage{makecell}

\usepackage{blindtext}

\usepackage{comment}

\usepackage{graphicx} %for including figure

\usepackage{subcaption} 
\usepackage{amssymb}
\usepackage{listings} % for insering code
\usepackage{verbatim}
\usepackage{booktabs}
\usepackage{float}

\usepackage{tabularx} % for shaping size of table
\usepackage{xcolor}

\usepackage{amsmath}

\usepackage{cite}
\usepackage{amsmath,amssymb,amsfonts}
\usepackage{algorithmic}
\usepackage{graphicx}
\usepackage{textcomp}

\title{Quality of Control Assessment for Tactile Internet based Cyber-Physical Systems}

\author
{
    \IEEEauthorblockN
    {
    Kurian Polachan\IEEEauthorrefmark{1},
    Joydeep Pal \IEEEauthorrefmark{2},
	Chandramani Singh\IEEEauthorrefmark{3}, and
	Prabhakar T V\IEEEauthorrefmark{4}
    }

    \IEEEauthorblockA
    {
        Indian Institute of Science, Bangalore, India\\
            \{\IEEEauthorrefmark{1}kurian, \IEEEauthorrefmark{2}joydeeppal, \IEEEauthorrefmark{3}chandra,
            \IEEEauthorrefmark{4}tvprabs\}@iisc.ac.in 
    }
    \thanks{An earlier version of this paper was presented at the IEEE International Conference on Sensing, Communication and Networking (IEEE SECON), Boston, USA, 2019 \cite{ref_polachanQoC}.}
}

\begin{document}

\onecolumn{
\section*{\textbf{\huge{IEEE Copyright Notice}}}

\huge{This work has been submitted to the IEEE for possible publication. Copyright may be transferred without notice, after which this version may no longer be accessible.} 
}
\newpage

\maketitle
%\input{myContents/chapWeeklyUpdate}
%\input{myContents/chapTestbedOnly}
%for iot journal

\begin{abstract}
We evolve a methodology and define a metric to evaluate Tactile Internet based Cyber-Physical Systems or Tactile Cyber-Physical Systems (TCPS). Towards this goal, we adopt the step response analysis, a well-known control-theoretic method. The adoption includes replacing the human operator (or master) with a controller with known characteristics and analyzing its response to slave side step disturbances. The resulting step response curves demonstrate that the \textit{Quality of Control} (QoC) metric is sensitive to control loop instabilities and serves as a good indicator of cybersickness experienced by human operators. 
%
%
%We evolve a methodology and define a metric to evaluate Tactile Cyber-Physical Systems (TCPS). Towards this goal, we adopt the step response analysis, a well-known control-theoretic method. The adoption includes replacing the human operator (or master) with a controller with known characteristics and analyzing its response to slave side step disturbances. The resulting step response curves demonstrate that the \textit{Quality of Control} (QoC) metric is sensitive to control loop instabilities and serves as a good indicator of cybersickness experienced by human operators. 
%%
We demonstrate the efficacy of the proposed methodology and metric through experiments on a TCPS testbed. The experiments include assessing the suitability of several access technologies, intercontinental links, network topologies, network traffic conditions and testbed configurations. Further, we validate our claim of using QoC to predict and quantify cybersickness through experiments on a teleoperation setup built using Mininet and VREP.

\end{abstract}

\section{Introduction}
% overview

\IEEEPARstart{A} teleoperation system consists of a human operator maneuvering a remote slave robot known as teleoperator over a network. The network transports the operator-side kinematic command signals to the teleoperator and provides the operator with audio, video, and/or haptic feedback from the remote-side. 
In existing teleoperation systems such as in telesurgery, due to significant end-end latency and reliability issues in both networking and non-networking components, the operator, i.e., the surgeon is forced to restrict his/her hand speed when he/she performs remote surgical procedures such as knot-typing, needle-passing, and/or suturing. %
This restriction in hand speed is required to avoid control loop instability and operator side cybersickness \cite{6755599,ref_tactileInternetITU,ref_Arjun}. Control loop instability occurs when the dynamics of the operator's hand actions exceed the response time of the teleoperation system.  Instability in control loops can cause slave robots to go out of synchronization with the operator's hand movements, which can lead to disastrous consequences in critical applications. Cybersickness occurs when there is a significant noticeable delay in the operator receiving feedback in response to his/her actions. Cybersickness can result in general discomforts such as eye strain, headache,  nausea, and fatigue and deter human operators from prolonged use of the teleoperation system \cite{ref_LaViola2000}.

More recently, researchers have envisioned combining tactile internet, a network architecture characterized by sub-millisecond latency and very high packet reliability ($99.999\%$), with advanced motion tracking hardware and high-speed robotics. This could help realizing teleoperation systems that do not impose restrictions on human hand speed \cite{6755599, ref_tactileInternetITU,ref_aijaz2018tactile}. We refer to these prospective teleoperation systems as Tactile Cyber-Physical Systems (TCPS). TCPS are envisioned to have applications in several domains where human skillset delivery is paramount, as in automotive, education, healthcare and VR/AR sectors. %Given the importance of TCPS, the emerging 5G technologies have put an enormous focus on tactile internet under the umbrella of URLLC.  
Realizing tactile internet where end-to-end latency and reliability are within the prescribed limits  is an important part. We think, however, that significant research concerning the design and evaluation is also an equally important part for building non-networking components such as hardware and its response, speed of algorithms and protocols that accompany a TCPS application.

In this paper, we formalize and investigate the evaluation part of TCPS research. A comprehensive evaluation method is necessary to assess and compare different TCPS implementations and to make objective decisions on the components to use for building TCPS. An evaluation method catering to these needs is still non-existent, and therefore, there is an immediate need to develop one \cite{ref_Steinbach, ref_Kuipers2017}. Towards designing an ideal evaluation method catering to TCPS needs, we propose the following objectives.

%This calls for the need to develop an evaluation method, to evaluate and compare different TCPS implementations and to make objective decisions on the components to use for building TCPSs \cite{ref_Kuipers2017,ref_Steinbach}. For the design of an ideal evaluation method catering to TCPS needs, we propose the following objectives:

%This calls for the need of a testbed to realize TCPS applications and an evaluation method to assess performance with the following objectives: 

\begin{itemize}

\item Cybersickness and control loop quality are the two main concerns in TCPS. To ensure that they do not affect TCPS implementations, we must capture these effects early in the TCPS design cycle.

%\item Cybersickness and control loop quality are the two main concerns in TCPS. To ensure that they do not affect TCPS implementations, our evaluation methodology must capture these effects early in the TCPS design cycle. %An evaluation method for TCPS should be capable of detecting both. 

\item The experiments constituting TCPS evaluation should be oblivious to the intended application. Additionally, the experimental results should be reproducible and free from subjective errors. 

\item The experimental results should be captured by a single metric that has a physical significance. This is necessary to judge the effect of different uncorrelated variables on the planned TCPS application. 

\end{itemize}

\subsection{Related Work}
%\todo[inline]{ -- Brief overview of existing evaluation methodologies and their issues}
Given the nature of TCPS with humans in the loop, an evaluation metric should have the ability to capture bidirectional end-to-end performance. However, in the current literature, TCPS evaluation is done using  Quality of Service (QoS) metrics or by subjective methods that employ human operators \cite{ref_Hamam, ref_Kusunose,ref_Jaafreh,ref_Tatematsu, ref_Sakr, ref_Correa, ref_Chaudhari}. These metrics use several QoS parameters such as latency, jitter, and packet drops, mostly limited to unidirectional end-to-end performance. In the context of TCPS, these metrics have limited utility in checking if an implementation can meet a given specification because these metrics do not relate the quality of the operator's experience with his/her dynamics. Certain latency and jitter may suit a TCPS with slow operator movement, but may not be adequate for another TCPS with faster operator movement. For instance, it is difficult to arrive at the right combination of these metrics that would result in stable control loops or reduced cybersickness. Thus it is difficult to use them for comparing the performance of different TCPS implementations.   
%--

Subjective evaluation methods assess responses from a large number of human operators to formulate a meaningful metric. This metric can be either subjective \cite{ref_Hamam, ref_Kusunose, ref_Tatematsu, ref_Jaafreh} or objective \cite{ref_Sakr,ref_Correa,ref_Chaudhari}. In either case,  subjective evaluation methods are time-consuming, prone to errors and application-specific, and fail to account for operator dynamics as discussed earlier. Importantly, subjective methods cannot truly assess the TCPS control loop quality. For instance, a human operator assigned to evaluate the TCPS may not recognize if the system implementation has low levels of instabilities such as minor overshoots, minor oscillations or minor steady-state errors due to his/her limited sensory perceptions. Low levels of instabilities may be acceptable for non-critical applications. However, for critical applications like telesurgery, even low levels of instabilities can cause injuries to the remote-side patients.\footnote{Note that when instability occurs, we will see its effects at both the operator side manipulator and the remote side robot.}

%A framework for an objective evaluation of TCPS control loop quality is discussed in 
%A framework for evaluation the control loop quality based on step response analysis method is discussed in .   

\subsection{Key Contributions}
\begin{itemize}

%\item \textit{Evaluation Model}:
%Objective evaluation of TCPS is required to be done in two steps. In the first step, one characterizes QoS of each of the TCPS control loops, i.e., of the control loops corresponding to different modalities such as haptic, audio, video and their combinations. In the second step, one assesses the quality of control in each of these loops. As assessing all the control loops requires a cumbersome infrastructure, we propose an alternative approach. We identify the critical control loop - the loop that has the most stringent round trip time (RTT) requirement and assesses only this loop. We argue that \textit{kinematic-haptic-video} is the critical loop (see Section~\ref{sec_evaluationModel}).

\item \textit{Evaluation Methodology}:
The evaluation method for TCPS has to be generic and objective. For this, we adopt the step response method, a classic control-theoretic method for analyzing the quality of closed-loop systems. In our work, we leverage this method for TCPS by framing an objective evaluation methodology around a typical TCPS implementation. This is accomplished by replacing the human operator with a controller with known characteristics and analyzing its response to slave side step changes. We had introduced this framework to characterize control loop quality of TCPS in \cite{ref_Polachan}. However, this work is the first attempt to use the step response method for evaluating TCPS.% (see Section~\ref{sec_EvaluationMethodology}).
%The evaluation method requires to be generic and objective. Taking this into account, we use the step response method, a classic control-theoretic method for analyzing the quality of closed-loop systems. In our work, we leverage this method for TCPS by framing an objective evaluation methodology around a typical TCPS implementation. This is accomplished by replacing the human operator with a controller with known characteristics and analyzing its response to slave side haptic sensor step changes. In literature, a similar framework to characterize control loop quality in TCPS exists \cite{ref_Polachan}. However, to the best of our knowledge, ours is the first attempt that uses step response method for evaluating TCPS (see Section~\ref{sec_EvaluationMethodology}).

%The evaluation method requires to be generic and objective. Taking this into account, we use the step response method, a classic control-theoretic method for analyzing the quality of closed-loop systems. In our work, we leverage this method for TCPS by framing an objective evaluation methodology around a typical TCPS implementation. This is accomplished by replacing the human operator with a controller with known characteristics and analyzing its response to slave side haptic sensor step changes. To the best of our knowledge ours is the first such attempt (see Section~\ref{sec_EvaluationMethodology}).

\item \textit{Quality of Control}:
In the quest for an index to grade TCPS, we propose the metric \textit{Quality of Control}(QoC). QoC is designed to capture the effect of different QoS parameters on the TCPS application. We can, therefore, use QoC as a one-stop metric to grade different TCPS implementations. Being based on the step response method, QoC inherently takes into account the stability of the TCPS control loop. Further, the relation we deduce between QoC and the maximum allowed hand speed of a human operator helps in the design of TCPS free from cybersickness. This relation also associates a physical significance to the metric. % (see Section~\ref{sec_qualityOfControl}).

%In the quest for an index to grade TCPS, we propose a metric called \textit{Quality of Control} (QoC) - it is designed to capture the effect of different QoS parameters on the TCPS application. This allows QoC to be used as a one-stop metric for grading different TCPS implementations. As QoC is based on the step response method, it inherently takes into account stability of TCPS control loop. Further, the relation we deduce between QoC and the permissible maximum speed of operator hand movements assists in the design of TCPS free from cybersickness. This relation also associates a physical significance to the metric (see Section~\ref{sec_qualityOfControl}).

%\item \textit{Performance Evaluation}
%The efficacy of the metric and methodology is demonstrated through experiments on a real TCPS testbed. This includes assessing the suitability of several access technologies, intercontinental links, transmission techniques and testbed configurations (see Section~\ref{sec_evaluation}).

\end{itemize}

\subsection{Outline}
%\todo[inline]{need to rewrite}
We organize this paper as follows: 
In Section~\ref{sec_evaluationModel}, we propose an evaluation model for TCPS. From the different control loops in TCPS, we identify the critical control loops to avoid cybersickness and control loop instability. We propose that the quality of these critical control loops is an indicator of TCPS quality.  
Section~\ref{sec_EvaluationMethodology} describes evaluation methodologies to assess the quality of the critical control loops. 
In Section~\ref{sec_qualityOfControl}, we propose the QoC metric and explain how to determine QoC from step response experiments. Section~\ref{sec_VmaxSection} describes the relation between QoC and the maximum allowed operator hand speed (to avoid cybersickness). 
In Section~\ref{sec_QoCInLiterature}, we detail several notions of QoC in real-time control literature and how they are different from our QoC.
In Section~\ref{sec_qocPerformanceCurve}, we introduce QoC performance curves and their use. We describe the evaluation of QoC in Section~\ref{sec_evaluation} and conclude the paper in Section~\ref{sec_conclusion}.

\section{Evaluation Model}
\label{sec_evaluationModel}
\begin{figure}[!tbp]
\centering
\includegraphics [width=0.9\linewidth]{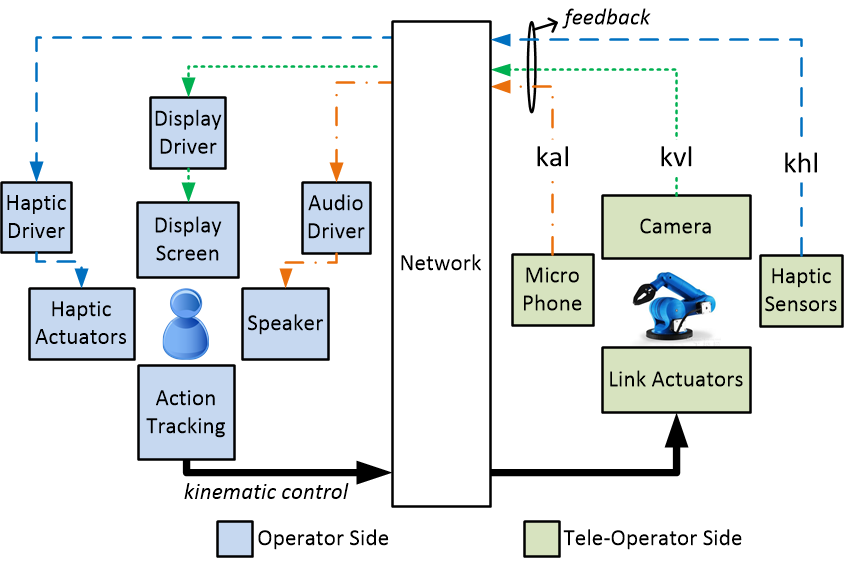}
\caption{Functional blocks and control loops in a typical TCPS. \textit{(1) kinematic-audio loop (kal)}, \textit{(2) kinematic-video loop (kvl)}, and \textit{(3) kinematic-haptic loop (khl)} control loops are shown. }
\label{fig_tcpsArchitecture}
\end{figure}

A typical TCPS can have multiple control loops, as shown in Figure~\ref{fig_tcpsArchitecture}. The control loops differ in their feedback modality. For instance, in the control loop \textit{kinematic-video}, the feedback is video. Similarly, for the control loops \textit{kinematic-audio} and \textit{kinematic-haptic}, the feedbacks are audio and haptic respectively. In all these cases, the feedback is in response to the same kinematic commands from the human operator. Here, the presence of a human operator warrants that these control loops adhere to the stringent QoS specifications, in particular concerning their Round Trip Times~(RTT). This is to save the operator from cybersickness and ensure control loop stability. 

Let us now discuss the RTT requirements for different TCPS control loops. We also identify the control loops that have the most stringent RTT requirement. We call these \textit{the critical control loops} and propose that the quality of these loop be used to benchmark the TCPS.
%\todo[inline]{black arrow in kal loop}

%\paragraph*{Discussion}
%In haptic feedback, haptic data from the tele-operator side is fed to the operator through haptic actuators. Instead, if the haptic data is displayed at the operator side screen we term it haptic-video feedback.

\subsection{Cybersickness}

Cybersickness to humans occurs as a result of conflict between different sensory systems. The conflict arises when different sensory systems perceive the occurrence of the same event at noticeably distinct times. In TCPS, cybersickness may result from the asynchronous arrival of different feedback modalities at the operator side in response to the same kinematic commands.  

We derive the maximum permissible feedback latencies (or maximum allowed RTTs) for different feedback modalities to avoid cybersickness.  These latencies determine the RTT specification of the corresponding TCPS control loops. Note that in a TCPS, delays incurred by both networking and non-networking components contribute to RTT of its control loops.

We begin by observing in Table~\ref{tbl_syncErrors}  the maximum permissible synchronization errors allowed for            
audio and haptic streams relative to video.\footnote{We consider a TCPS use case where audio and video are streamed independently of each other, unlike the case with streaming methods like MPEG.} Note that the maximum permissible synchronization errors can be either positive or negative depending on whether the  audio and haptic streams arrive after or ahead of the video. However, for TCPS, these streams arriving ahead of the video are not a cause of concern. Hence, we do not account for negative values of synchronization errors.

%%% table control loop specifications
\begin{table}[!htbp]
\centering
\caption{Maximum permissible $+ve$ synchronization errors of different media streams relative to video \cite{ref_Levitin, ref_Silva}.}
\centering
\begin{tabular}{c c}
\toprule
\makecell{Media \\ Streams} &  \makecell{Max Permissible $+ve$ Sync Error \\ Relative to Video}  \\ 
\midrule
audio  &  $45ms$    \\ 
                      
haptic       &$125ms$  \\

\bottomrule
\end{tabular}
\label{tbl_syncErrors}
\end{table}

We now derive the maximum permissible feedback latencies (or, maximum allowed RTTs of the corresponding control loops) using these maximum permissible synchronization errors. We show the results in Table~\ref{tbl_loopSpec}. We assume that the display screen at the operator end shows the actual size of the remote field. 

%%% table control loop specifications
\begin{table}[!htbp]
\centering
\caption{RTT specification for TCPS control loops.}
\centering
%\begin{tabular}{p{0.3\linewidth}  p{0.3\linewidth} p{0.3\linewidth}}
\begin{tabular} {c c c}
\toprule
\makecell{Feedback Modality} &  \makecell{Control Loop Specification } \\ 
\midrule

video       &$RTT_{kvl} \leq 1ms$ \\
audio         &$RTT_{kal} \leq 46ms$                          \\ 
haptic        &$RTT_{khl} \leq 126ms$                          \\ 

\bottomrule
\end{tabular}
\label{tbl_loopSpec}
\end{table}

\begin{figure}[!tbp]
\centering
\includegraphics [width=0.5\linewidth]{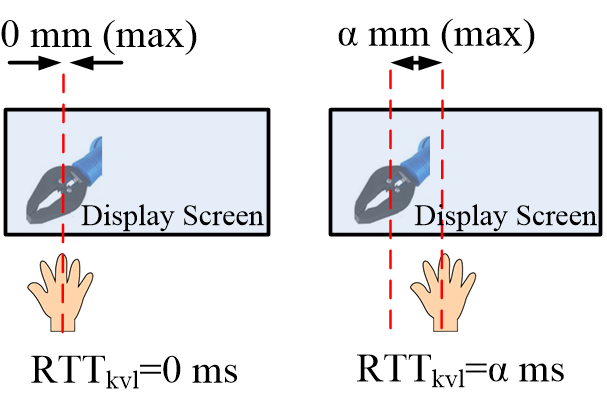}
\caption{The error between the operator's hand and robotic arm displayed on the screen for a hand speed of $1mm/ms$.} % The error is $\alpha\,mm$ when RTT of the corresponding control loop is $\alpha\, ms$.}
\label{fig_tcpsRttEffect}
\end{figure}

When there is a delay in video feedback, the operator will see a lack of synchronization between the movement of his/her hand and the remote robotic arm displayed on the screen.  A typical human operator can move his/her hands at a maximum speed of $1mm/ms$. Moreover, he/she can visually distinguish differences greater than or equal to $1mm$. Therefore, to avoid $1mm$ or larger differences in positions of the operator's hand and the robotic arm displayed on the screen, the maximum permissible delay for the video feedback is required to be $1ms$. This restricts the RTT of \textit{kinematic-video} loop, $RTT_{kvl}$, to be less than or equal to $1ms$ \cite{6755599} (See  Figure~\ref{fig_tcpsRttEffect}~(a)).
\footnote{Observe that if we restrict the human operator's maximum hand speed to $v$ m/s, $RTT_{kvl}$ can be allowed to be up to $1/v~ms$. Furthermore, if the display screen at the operator's end shows a zoomed out picture of the remote field (say a zoom factor $\alpha < 1$), then even with a maximum hand speed of $1 m/s$ and RTT of 1ms, the synchronization error seen by the operator will be less than $\alpha~mm$. In this case, $RTT_{kvl}$ could be allowed to be larger.    }

When the operator moves his/her hand, he/she may expect to hear the sound from the remote environment (e.g. sound of the moving motor joints of the remote side robot or sound of the robot hitting a target) within a specific time limit. If the delay in audio feedback is more than this limit, the operator will find a lack of synchronization between the movement of his/her hand and the audio response. This will also result in cybersickness. We compute the maximum permissible latency for the audio feedback by adding the maximum permissible latency for the video feedback and the maximum permissible $+ve$ synchronization error of audio relative to the video stream. We thus get the maximum permissible $RTT_{kal}$ to be $46ms$.

Following similar reasoning, we find the maximum permissible latency of the haptic feedback to $126ms$. This restricts $RTT_{khl}$ to $126ms$.

\subsection{Control Loop Instability}
As already explained, to avoid cybersickness, the RTT requirement on the \textit{kinematic-haptic} loop is much higher than the requirement on the \textit{kinematic-video} loop. Thus, towards avoiding cybersickness, the \textit{kinematic-haptic} loop is not critical. However, in TCPS where haptic feedback exists in the form of kinesthetic feedback\footnote{Haptic feedback is of two types, kinesthetic and tactile. Kinesthetic feedback provides information about the stiffness of materials, while tactile feedback provides information concerning the texture and friction of material surfaces. We sense kinesthetic feedback through our muscles, joints and tendons while we sense the tactile feedback through the mechanoreceptors of our skin \cite{2452454}. 
}, RTT of the \textit{kinematic-haptic} loop, $RTT_{khl}$, becomes critical as follows.

In teleoperation systems with kinesthetic feedback, the operator experiences the haptic property (e.g. stiffness) of objects in the remote environment by commanding the teleoperator to tap the surface of these objects at different taping velocities and sensing the feedback forces through the operator side manipulator \cite{ref_Steinbach_HapticCodecs}. For instance,  to differentiate a hard material from a soft material, the operator taps the surface at a higher velocity. At a higher tapping velocity, the force feedback felt by the operator from a hard material will be higher in comparison to a soft material. In TCPS with kinesthetic feedback, to support for the maximum tapping velocity of a typical human operator and thereby enable the operator to differentiate materials of wide stiffness range, the RTT of the \textit{kinematic-haptic} loop, $RTT_{khl}$ is required to be less than $1ms$~\cite{ref_Steinbach}. RTT higher than $1ms$ implies that the operator has to restrict his tapping velocity and thus limit his/her ability to differentiate materials of higher stiffness. However, if tapping velocity is not restricted, depending on the stiffness of the material, the \textit{kinematic-haptic} loop may go unstable. This results in high-frequency oscillations in the operator side manipulator and remote side robot. 

\begin{figure}[!htbp]
\centering
\includegraphics [width=0.95\linewidth]{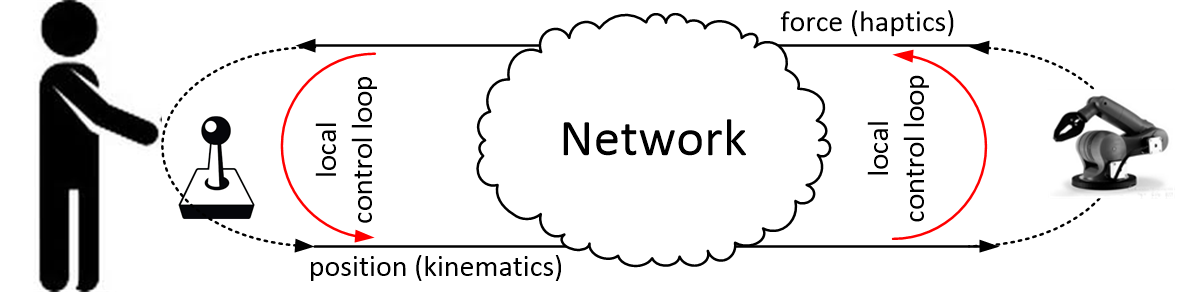}
\caption{Presence of fast-acting local control loops in a TCPS with kinesthetic feedback \cite{ref_SteinbachPPTHapticComm}. } 
\label{fig_tcpsWithKinestheticFeedback}
\end{figure}

Unlike in \textit{kinematic-audio} and \textit{kinematic-video} loops, control loop instability can occur in the \textit{kinematic-haptic} loop due to the existence of fast-acting local control loops at the operator and teleoperator sides.% meant for exchanging energy between subsystems \cite{ref_Steinbach_hapticCommunications}.
These local control loops close the global \textit{kinematic-haptic} control loop detouring the human operator (see~Figure~\ref{fig_tcpsWithKinestheticFeedback}).  In regular operation, the local control loops are meant to synchronize both the position and force variables between the operator and the teleoperator sides. However, at higher RTTs, higher tapping velocities and in the presence of stiff materials, they assist in the generation of positive feedback destabilizing the global control loop \cite{ref_Steinbach_hapticCommunications, ref_Steinbach_HapticCodecs}.  

\subsection{Control Loop for Evaluation}
We propose to benchmark a TCPS by evaluating the quality of its critical control loop, the one with the most stringent RTT requirement. We have seen that both \textit{kinematic-video} and \textit{kinematic-haptic} loops are critical loops for avoiding cybersickness and control loop instability, respectively. We propose evaluating one of these loops, depending on which loop is critical for a given TCPS application. In case, if both these loops are equally critical, we propose to evaluate both.

%\todo[inline]{need to rewrite}

In Table~\ref{tbl_controlLoopCritical}, we list these possible cases and corresponding possible TCPS scenarios. The occurrence of cybersickness is prominent, i.e., $RTT_{kvl}$ is more critical when the operator hand speed is high. The occurrence of control loop instability is prominent, i.e., $RTT_{khl}$ is more critical when the stiffness of the remote-side object is high for medium and high hand speeds. 
\begin{table}[!htbp]
\centering
\caption{Criticality of control loops for different TCPS scenario's.
}

\begin{tabular}{@{}c|c|c|l@{}}
\toprule
\textbf{}       & \textbf{RTT$_{kvl}$} & \textbf{RTT$_{khl}$} & \multicolumn{1}{c}{\textbf{Sample Scenario}}                                        \\ \midrule
\textbf{Case-1} & \checkmark          & \checkmark
          & \begin{tabular}[c]{@{}l@{}}Hand Speed = High\\ Object Stiffness = High\end{tabular} \\ \midrule
\textbf{Case-2} & \checkmark          & -          & \begin{tabular}[c]{@{}l@{}}Hand Speed = High\\ Object Stiffness = Low\end{tabular} \\ \midrule
\textbf{Case-3} & -          & \checkmark          & \begin{tabular}[c]{@{}l@{}}Hand Speed = Medium\\ Object Stiffness = High\end{tabular} \\ \bottomrule
\end{tabular}
\label{tbl_controlLoopCritical}
\end{table}

\section{Evaluation Methodology}
\label{sec_EvaluationMethodology}
For evaluating TCPS, we use the step response method, a classic control-theoretic method for assessing the quality of closed-loop systems. We leverage this method for TCPS by replacing the human operator with a known Proportional Integral (PI) controller and analyzing its response to the slave side step changes. In this section, we first describe the components of the TCPS testbed implemented in our lab \cite{ref_Polachan}. We then describe how to conduct step response experiments using this testbed in both haptic and non-haptic settings to evaluate \textit{kinematic-haptic} and \textit{kinematic-video} loops, respectively.

%For evaluating TCPS, we use the step response method, a classic control-theoretic method for assessing the quality of closed-loop systems. We leverage this method for TCPS by replacing the human operator with a known Proportional Integral (PI) controller and analyzing its response to the slave side step changes. In this section, we first describe the components of the TCPS testbed implemented in our lab \cite{ref_Polachan}. We then explain how to conduct the step response experiments using this testbed on the \textit{kinematic-haptic} loop.

\subsection{Component Level Design of the Testbed}

\begin{figure}[!htbp]
\centering
\includegraphics [width=0.95\linewidth]{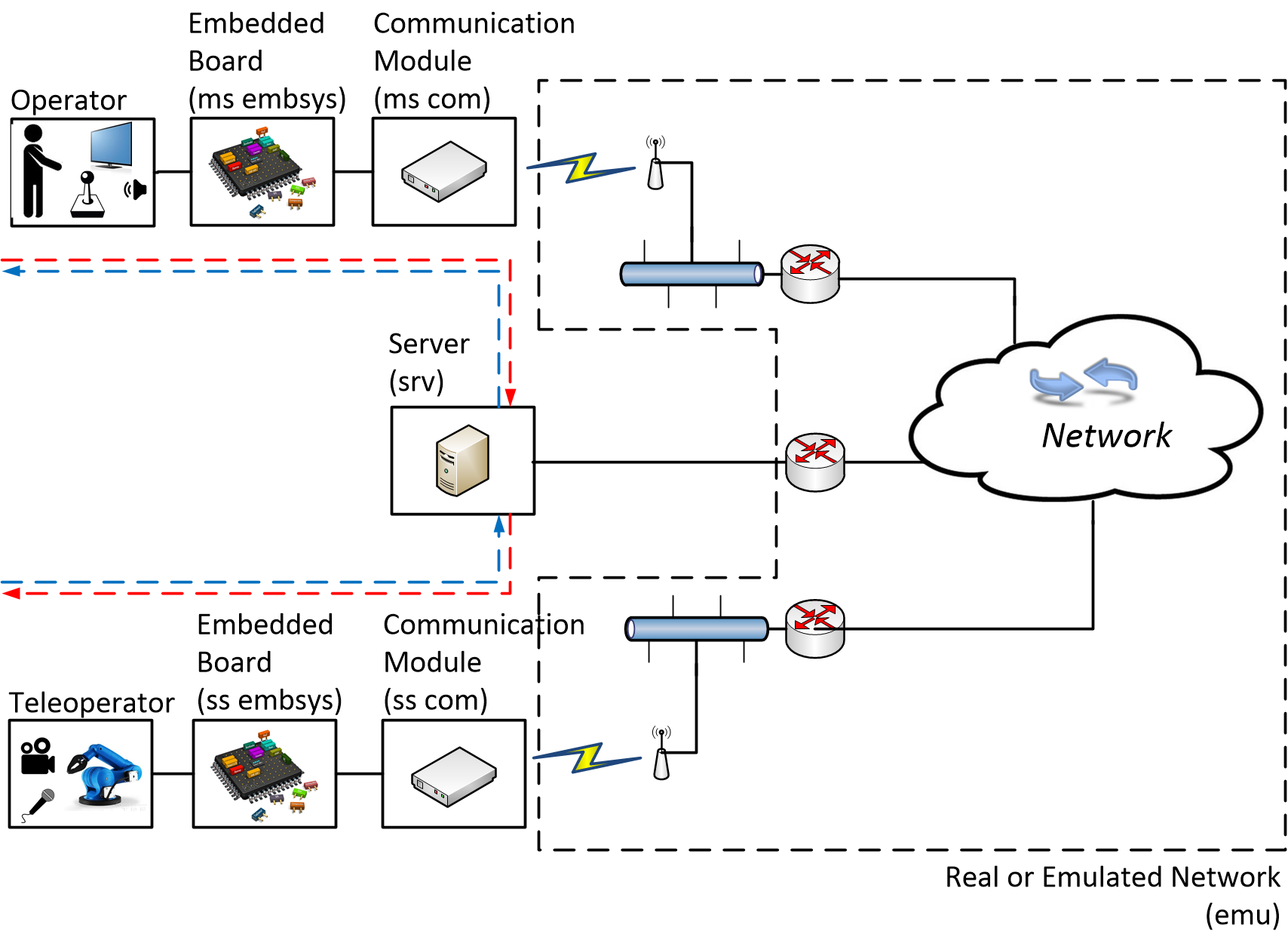}
\caption{Component level design of the testbed with forward and
backward data flow paths marked \cite{ref_Polachan}.}
\label{fig_testBed}
\end{figure}

The component level design of the testbed is shown in Figure~\ref{fig_testBed}.
\begin{itemize}

\item \textit{operator} is the human operator or an embedded controller; \textit{tele-operator} is the remote side slave device being controlled.

\item \textit{ms embsys} is the master side embedded system which houses sensors, actuators, and algorithms to capture the kinematic motions of the operator, to display audio and video, and to apply haptic feedback to the operator.
%
%\item 
\textit{ss~embsys} is the slave side embedded system that houses sensors, actuators, and algorithms for driving the slave side robot and for capturing the remote audio, video, and haptic signal. 
%audio/video sensors and tactile sensors to capture the haptic information.

\item \textit{ms com} is the master side communication component that connects \textit{ms~embsys} to the network; \textit{ss com} is the slave side counterpart which connects \textit{ss embsys} to the network.

%\item \textit{srv} is the computer for various computations in the testbed. It executes relevant algorithms in the forward and the backward data paths. 
\item \textit{srv} is the computer for offloading various TCPS algorithms in sensing, actuation, coding, and compression. %\emph{srv} can give insights into the real processing overheads faced by embedded computers of TCPS.

\item \textit{emu} is the optional emulator used to interconnect the testbed components. The emulator replaces a real physical network. We use ns-3 and Mininet to build this emulator \cite{ref_ns3, ref_mininet}. 

\end{itemize}

%\section{Evaluation Methodology}
%\label{sec_EvaluationMethodology}

%In this section, we explain the experimental setup we use to evaluate a TCPS application. We 

%\section{Evaluation Methodology}
%\label{sec_EvaluationMethodology}

%In this section, we explain the experimental setup we use to evaluate a TCPS application. We 

\subsection{Evaluation Framework in a Haptic Setting}
\label{sec_evaluationFramework}
\begin{figure}[!htbp]
\centering
\includegraphics [width=.85\linewidth]{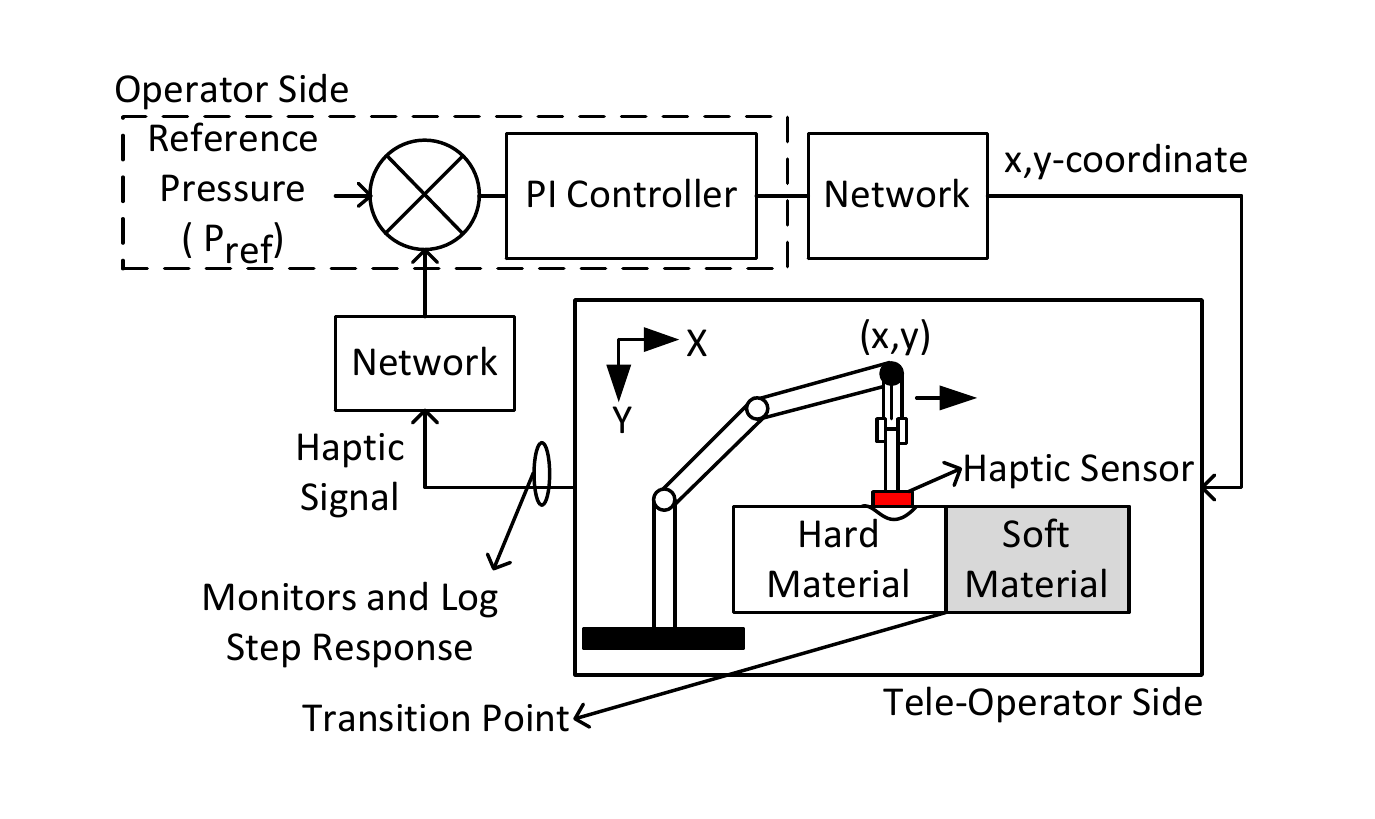}
\caption{Evaluation framework}
\label{fig_controlModel}
\end{figure}

Evaluating the \textit{kinematic-haptic} loop helps in quantifying control loop instability in a TCPS. For evaluating the \textit{kinematic-haptic} loop, we propose the evaluation framework in Figure~\ref{fig_controlModel}. To develop this, we use the testbed in Figure~\ref{fig_testBed} and replace the operator with a PI controller.  At the teleoperator side of the testbed, we place a robotic arm with a haptic sensor mounted to its end effector. The haptic sensor detects pressure when the end effector of the robotic arm comes in contact with the material surface. In the step response experiment, we use the PI controller to move the robotic arm along the X-axis and to apply constant pressure of $P_{ref}$ on the material surface along its Y-axis. When the robotic arm crosses the transition point separating the hard and the soft materials, the pressure abruptly drops from $P_{ref}$ to $P_{ref}/k_2$, simulating a step-change in the haptic domain. Here $k_2$ is a constant and is dependent on the characteristics of the materials in use.

%The evaluation framework is shown in Figure~\ref{fig_controlModel}. To develop this, we use the testbed in Figure~\ref{fig_testBed} and replace the operator with a PI controller.  At the teleoperator side of the testbed, we place a robotic arm with a haptic sensor mounted to its end effector. The haptic sensor detects pressure when the end effector of the robotic arm comes in contact with the material surface. In the step response experiment, we use the PI controller to move the robotic arm along the X-axis and to apply constant pressure of $P_{ref}$ on the material surface along its Y-axis. When the robotic arm crosses the transition point separating the hard and the soft materials, the pressure abruptly drops from $P_{ref}$ to $P_{ref}/k_2$, simulating a step-change in the haptic domain. Here $k_2$ is a constant and is dependent on the characteristics of the materials in use.
%
The haptic sensor detects this change in pressure and communicates the new pressure to the operator side PI controller. The PI controller uses this data to compute a new $y$-coordinate of the robotic arm and communicates it back to the teleoperator side to take action. The intent here is to adjust $y$ to increase the applied pressure and correct the haptic step change. 

\begin{figure}[!htbp]
\centering
\includegraphics [width= 0.75\linewidth]{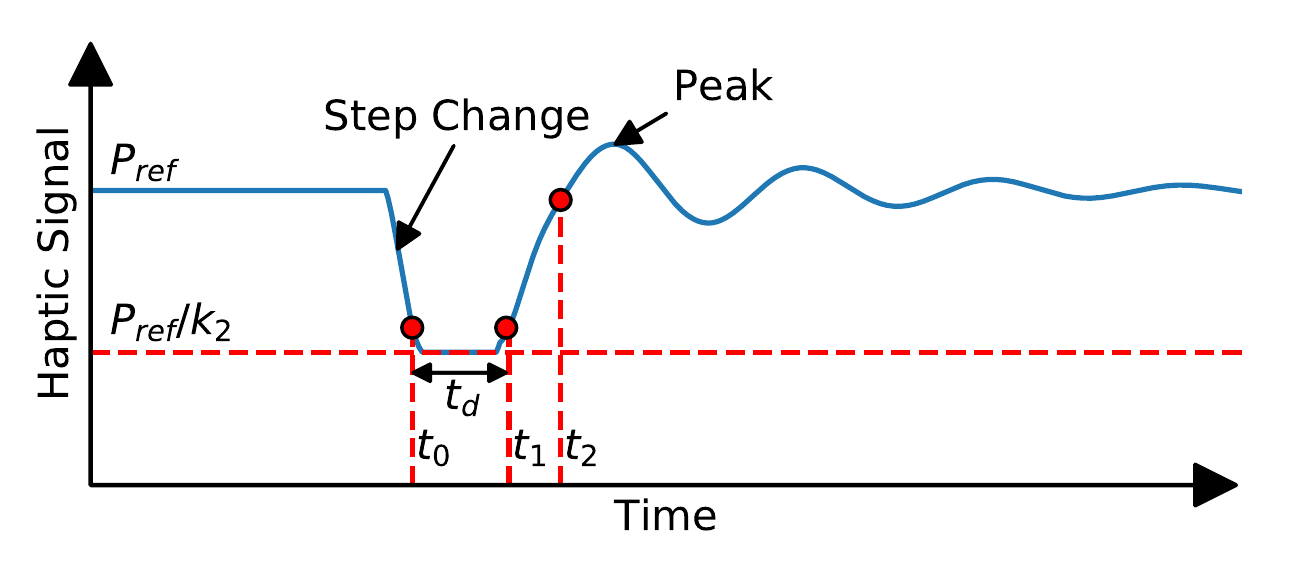}
\caption{Sample haptic step response curve. Haptic signal at time instants $t_0$, $t_1$, and $t_2$ are marked using red dots. 
}
\label{fig_evaluationStepInput}
\end{figure}

Figure~\ref{fig_evaluationStepInput} shows a sample haptic step response curve expected from the experiment. The curve resembles the classic step response curve in control system with its quality determined by certain rise time, overshoot, undershoot, settling time, and steady-state error. The quality of this curve directly indicates the quality of the \textit{kinematic-haptic} loop, which in turn indicates the quality of the TCPS under test. 
In our work, we define \textit{rise time}  as $t_r = t_2 - t_0 = t_d + t_2-t_1$ where,
\begin{enumerate}
    \item $t_0$ is the time at which the pressure drops to $P_{ref}/k_2 + 0.1(P_{ref} - P_{ref}/k_2)$ (due to step change).
    \item $t_1$ is the time at which the pressure rises back to $P_{ref}/k_2 + 0.1(P_{ref} - P_{ref}/k_2)$.  
    \item $t_2$ is the time at which the step response reaches $P_{ref}/k_2 + 0.9(P_{ref} - P_{ref}/k_2)$.  
\end{enumerate}

%the time taken by the step response to rise from $(P_{ref}/k_2+0.1(P_{ref}-P_{ref}/k_2))$ at $t_0$ to $(P_{ref}/k_2+0.9(P_{ref}-P_{ref}/k_2))$ at $t_1$ plus $t_d$, the delay in controller response. 
We define \textit{overshoot} as the peak percentage fluctuation in the step response relative to $(P_{ref}-P_{ref}/k_2)$. Both the rise time and overshoot are affected by the characteristics of the TCPS components and the network.

%In our work, we define \textit{rise time}, $t_r$, as the time taken by the step response to rise from $(P_{ref}/k_2+0.1(P_{ref}-P_{ref}/k_2))$ at $t_0$ to $(P_{ref}/k_2+0.9(P_{ref}-P_{ref}/k_2))$ at $t_1$ plus $t_d$, the delay in controller response. We define \textit{overshoot} as the peak percentage fluctuation in the step response relative to $(P_{ref}-P_{ref}/k_2)$. Both the rise time and overshoot are affected by the characteristics of the TCPS components and the network.

\paragraph{Operator Side Implementation}
We describe the implementation of the PI controller using the code snippet in Figure~\ref{fig_codeOperatorSide}. Here in every loop, the $x$ coordinate of the robotic arm is incremented by $1cm$. The parameter $\Delta$ is the loop wait time used for tuning the responsiveness of the controller and also the stability of the system. A controller using smaller $\Delta$ can potentially respond faster to the haptic signal, and therefore lead to smaller rise times. However, setting a very small $\Delta$ can potentially result in overshoots and oscillations in the step response curve. In Figure~\ref{fig_codeOperatorSide}, $k_p$ is the PI controller constant.

%We first initialize the robot end effector coordinates. We choose the initial values such that the robot end effector is at X=0units (see Figure~\ref{fig_evaluationSetup} and the robot is applying a pressure of $refP$ on the surface of the hard material. We then move the robot 1units towards the right in every loop. In every loop, we check for haptic signal from tele-operator side. If we receive data, we update the variable $p$ and find error in pressure. If we do not receive any haptic signal we use the old $p$ value to compute the error. We try to nullify this error by adjusting the $Y$ coordinate of the robot end effector. To determine $Y$ we use the PI controller equation with controller constant $Kp$. In the code, we add a loop wait time of $\Delta$ seconds, this is to adjust the responsiveness of the controller. If $\Delta$ is low, the controller acts faster else it acts slower to the haptic signal it receive. See Figure~\ref{fig_codeOperatorSide} for the algorithm.

\begin{figure}[!htp]
\begin{lstlisting} [basicstyle=\ttfamily\normalsize, morekeywords={function, while, if, else, wait, OperatorSide},frame=tb, escapeinside={(*}{*)},mathescape=true]
OperatorSide ():
  initialize coordinates [$x=0$,$y=0$]
  while ($x < Xcm$):	
    send robot coordinates $x$ and $y$	
    wait for $\Delta$ seconds
    if (haptic signal available):
	receive and store in variable $P$		
    else:
    	store old haptic signal in $P$
    find $error$ = $P_{ref}$ - $P$
    compute $y = y + k_p \times error$
    increment $x$ by $1cm$    
\end{lstlisting}
\caption{PI controller implementation.} 
\label{fig_codeOperatorSide}
\end{figure}

%\todo[inline]{Replace symbol N with delta}
%\todo[inline]{Give units for K1, K2, refP, N and Kp}

\paragraph{Tele-Operator Side Implementation}

In the evaluation framework in Figure~\ref{fig_controlModel}, we use a robot and a force sensor to simulate and measure the haptic step change. This implementation is useful if, in the TCPS evaluation, we desire to account for the characteristics of the robot and the force sensor. Often we are not interested in accounting for the robot's or sensor's physical limitations (e.g. when we want to evaluate the TCPS communication network). In such cases, we replace the teleoperator side with a code snippet, as shown in Figure~\ref{fig_codeTeleOperatorSide2}. In this code snippet, we have used $k_1$ as the conversion constant to translate $y$ to pressure. The change in pressure from $k_1y$ to $k_1y/k_2$ is simulated at $x=X/2\,cm$ (we choose $X=100$ in our experiments). 

%% algorithm
\begin{figure}[!tbp]
\begin{lstlisting} [basicstyle=\ttfamily\normalsize, morekeywords={TeleOperatorSide, while, if, else, wait},frame=tb,escapeinside={(*}{*)},mathescape=true]
TeleOperatorSide ():
 while (true):	
  wait to receive coordinates $x$ and $y$
  if ($x < X/2$):
   set haptic signal to $k_1y$
  if ($x \geq X/2$):
   set haptic signal to $k_1y/k_2$   
  send haptic signal to operator side
  log haptic signal to file
\end{lstlisting}
%\caption{Code snippet used for simulating haptic step change at the tele-operator side.}
\caption{Simulating haptic step change at the tele-operator side.} 
\label{fig_codeTeleOperatorSide2}
\end{figure}

\subsection{Evaluation Framework in a Non-Haptic Setting}

\label{sec_evaluationFrameworkNonHapticSetting}

\begin{figure}[!htbp]
\centering
\includegraphics [width=0.8\linewidth]{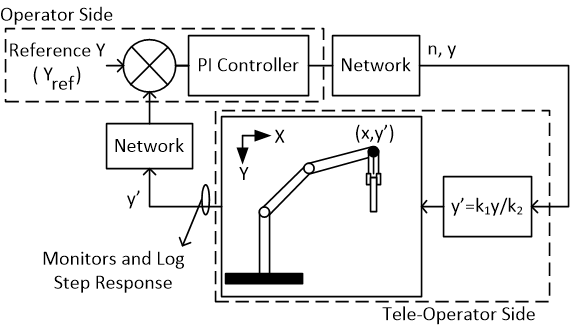}
\caption{Evaluation framework for a non-haptic setting.}
\label{fig_controlModelNonHaptic}
\end{figure}

Evaluating \textit{kinematic-video} loop of a TCPS helps in determining $V_{max}$, the maximum operator hand speed the TCPS can support to avoid cybersickness. For evaluating the \textit{kinematic-video} loop, we cannot use the evaluation framework in Figure~\ref{fig_controlModel} as it simulates step change in the non-existing haptic domain.  In Figure~\ref{fig_controlModelNonHaptic}, we propose a modified evaluation framework. We use the testbed in Figure~\ref{fig_testBed} and replace the operator using a PI controller with controller constant, $k_p$, and loop wait time parameter $\Delta$. At the teleoperator side of the testbed, we place a robotic arm with a video camera to detect the $y$-coordinate of the robot end-effector. 

In the step response experiment, we use the PI controller to always maintain a constant y-coordinate, $Y_{ref}$, for the robotic arm. The PI controller in every control loop (i.e., after every $\Delta$ interval of time), increments, and send an epoch variable, $n$, to the teleoperator side. The PI controller initializes $n$ to $1$ at the start of the experiment. At the teleoperator side, $n$ is used to set the value of $k_2$. The teleoperator initializes $k_2$ to $1$ at the start of the experiment and is set to a higher value when $n$ is $\geq50$. This simulates a step-change in the robot $y$-coordinate, $y'$, at $n=50$. The video camera at the teleoperator side detects the change in $y$-coordinate of the robot and communicates this new $y$-coordinate to the operator side PI controller. The PI controller uses this data to compute $y$, a new $y$-coordinate for the robotic arm and communicates it back to the teleoperator side to take action. The intent here is to adjust $y$ to correct the step-change in $y'$. The parameters of the resultant step response curve, i.e., the plot of $y'$, is similar in shape to Figure~\ref{fig_evaluationStepInput} and is used to evaluate the quality of the \textit{kinematic-video} loop and thus the TCPS under test.

As for the evaluation framework in Figure~\ref{fig_controlModel}, here also, we can replace the robot and the video camera with a code snippet if accounting the overhead of these components is not desired in evaluation.

\paragraph*{Discussion}  
In the evaluation framework, we discount the characteristics introduced by the display driver block. This is because display driver overheads are generally deterministic and known apriori, and their impact can be theoretically determined. 

\section{Quality of Control}
\label{sec_qualityOfControl}
In this section, we first describe how to arrive at the optimum value of $\Delta$, denoted as $\Delta_{opt}$. We then describe how to determine the parameters $P_{ref}$, $k_1$, $k_2$ and $k_p$ of the evaluation framework. We then describe QoC, the evaluation metric we propose for TCPS. For descriptions, we consider step response curves from the evaluation framework proposed for evaluating the \textit{kinematic-haptic} loop. Note that the methods we describe here are also valid for the evaluation framework proposed for a non-haptic setting, i.e., for evaluating the \textit{kinematic-video} loop. In the following, we write rise time as $t_r(\Delta)$ to show its dependence on $\Delta$. 

\subsection{Determining $\Delta_{opt}$}
The step response curve extracted from the evaluation setup depends on the characteristics of the TCPS under test and also on $\Delta$. For any given TCPS, different values of $\Delta$ lead to different step response curves. If $\Delta$ is very small, the resulting step response curve can have oscillations and overshoots. As we increase $\Delta$, the controller's response gets slower, and hence oscillations and overshoots reduce, and the rise time increases. \textit{We define a good step response curve to be a response curve in which the overshoots and steady-state error are within prescribed limits - we set these limits to $20\%$ and $10\%$ respectively in our work}\footnote{In our work, we set limits only on overshoot and steady-state error to identify good step response curves. This is to simplify the classification of step response curves. In practice, we advise setting limits on all step response curve parameters including undershoot and settling time.}. We want to achieve the fastest good response curve, i.e., the good response curve with the least rising time. Accordingly, we should set $\Delta$ to the least value that results in a good response curve; we refer to this value as $\Delta_{opt}$.  Alternatively stated, $\Delta = \Delta_{opt}$ results in a response curve that has the least rise time among all the response curves with overshoots and steady-state errors within $20\%$ and $10\%$ respectively.  In the following, we refer to this good response curve as a $\Delta_{opt}$-curve.

\begin{figure}[!htbp]
\centering
\includegraphics [width=0.80\linewidth]{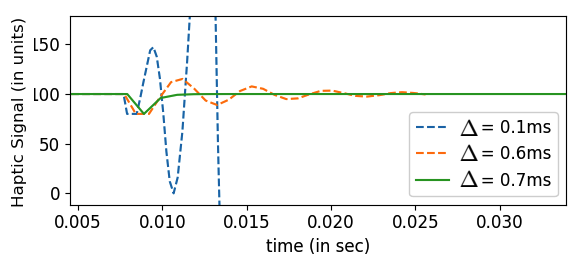}
\caption{Step response curves of a sample TCPS for a few values of $\Delta$. Here $P_{ref} = 100$ and $P_{ref}/k_2 = 80$.}
\label{fig_findingNopt}
\end{figure}

We determine $\Delta_{opt}$ through experiment. For the TCPS under test, Figure~\ref{fig_findingNopt} shows the step response curves for a few values of $\Delta$. Observe that, with $20\%$ and $10\%$ limits, the overshoots and steady-state error for good response curves should be less than $4units$ and $2units$ respectively. In particular, peaks of good response curves should be less than $104units$. $\Delta=0.1ms$ and $\Delta=0.6ms$ yield response curves with peaks exceeding $104units$. On the other hand, $\Delta=0.7ms$ yields the good step response curve with the least rise time. We thus determine $\Delta_{opt}=0.7ms$. Note that a smaller $\Delta_{opt}$ implies a potentially smaller $t_r$ and hence a better quality TCPS.

%
%The $\Delta_{opt}$ is determined through experiments. For a sample TCPS under test,  Figure~\ref{fig_findingNopt} shows the step response curves for different values of $\Delta$. For the experiments $P_{ref}$ is set to $100$ units and $k_2$ to $1.25$. For these settings to restrict the overshoot to $20\%$, as per the definition of overshoot in Section~\ref{sec_evaluationFramework}  the peaks of the step responses curves should not exceed $104$ units. We find that $\Delta=0.1ms$ and $\Delta=0.6ms$ do not yield a good step response curve as their peaks are above $104 units$. We find that $\Delta=0.7ms$ yields a good step response curve. Hence we determine $\Delta_{opt}$ as $0.7ms$. Note that a smaller $\Delta_{opt}$ indicates a better quality TCPS. 
%
\subsection{Design of Evaluation Parameters}
\label{sec_designOfEvalutionParameter}
\begin{figure}[!htbp]
\centering
\includegraphics [width=0.820\linewidth]
{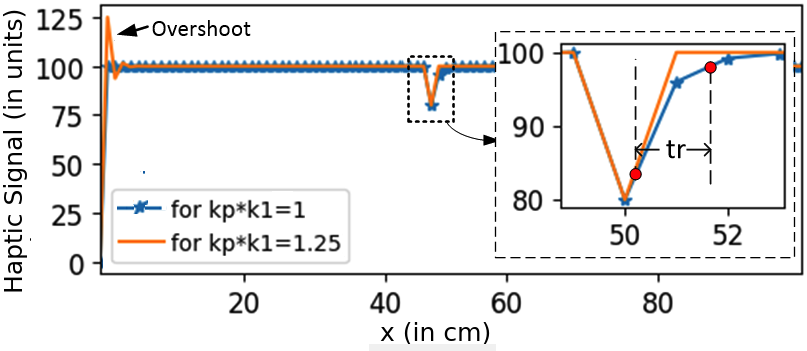}
\caption{Haptic step response curve for different $k_pk_1$ settings for a TCPS with zero packet drops and $RTT_{\max} < \Delta$. For $k_pk_1$ = 1.25, we see overshoot at the start of the step response experiment.}
%\caption{Magnified step input portion from Figure~\ref{fig_ControlParameterSettings}}
\label{fig_ControlParameterSettingsZoomIn}
\end{figure}

%For a TCPS, RTT may vary with time in which case we consider the maximum of these RTT's as $RTT_{max}$. 
For a TCPS with zero packet loss and $RTT < \Delta$, we can deduce a difference equation from the basic PI controller equation as follows. Let $y_l$ be the current $y$-coordinate of the robot and $y_{l+1}$ be its next $y$-coordinate determined by the PI controller. Then, for stability and fast convergence of this difference equation, the roots of its characteristic equation in the $Z$ domain, $z - \left(1 - k_pk_1/k_2 \right) = 0$, should be within the unit circle and close to the origin. 
\begin{align*}
y_{l+1} &= y_l + k_p \times error  \\
 &= y_l + k_p (P_{ref} - P) \\ 
 &= y_l + k_p \left(P_{ref} - \dfrac{k_1y_l}{k_2}\right) \\
 &= y_l\left(1 - \dfrac{k_pk_1}{k_2}\right) + k_pP_{ref} 
\end{align*}
We see from the characteristic equation that $P_{ref}$ does not influence the stability or the convergence speed of the difference equation. Thus we can choose any value for $P_{ref}$. For ease of design we fix $P_{ref}=100$ units.  The parameter $k_2$ allows realizing the haptic step. We want to set $k_2$ to restrict the step signal to $20\%$ of the $P_{ref}$. This is necessary to maintain the operation of the slave device around its operating point. We thus set $k_2=1.25$.  Furthermore, we could set $k_p k_1= k_2 = 1.25$ so that the root of the characteristic equation would be at the origin of the Z-plane. This would ensure that the configuration of the evaluation setup does not mask the characteristics of the TCPS under test, and we also achieve the fastest possible step response.  However, this results in an overshoot at the start of the step response experiment (See Figure~\ref{fig_ControlParameterSettingsZoomIn}).  This is because, initially, when $x < 50units$ the effective value of $k_2$ is $1$ which results in the root of the characteristic equation to be in the left half of the Z-plane.  In order to ensure that the root of the characteristic equation is positive all through, we set $k_p k_1 = 1$. In particular, we choose $k_1 = k_p = 1$. 

\begin{figure}[!htbp]
\centering

\includegraphics [width=0.85\linewidth]{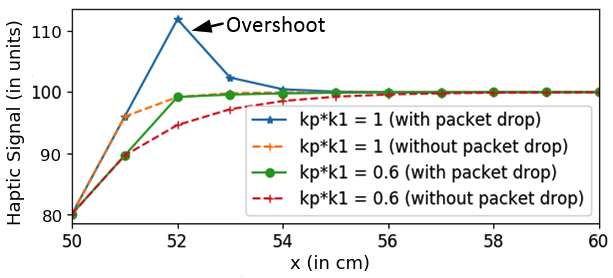}
\caption{Haptic step response curve (zoomed-in) for different $k_pk_1$ settings; with and without packet drops.}
\label{fig_ControlParameterSettingsPacketLoss}
\end{figure}

We do not recommend setting $k_p k_1 < 1$. A lower $k_p k_1$ can increase the response time and can impact the ability of the evaluation methodology to detect packet drops or latency issues commonly observed with the TCPS communication networks. We demonstrate this in Figure~\ref{fig_ControlParameterSettingsPacketLoss}. We simulate packet drops in the communication network of a TCPS under test. We see that packet drops cause an overshoot when $k_pk_1 = 1$ but do not when $k_pk_1 = 0.6$. Hence we can detect packet drops from overshoot in the former case but cannot detect these in the latter case.

\subsection{Design of Evaluation Metric}
\label{sec_designOfEvalutionMetric}
To evaluate TCPS, we propose a metric that is an indicator of TCPS control loop quality. We call this metric \textit{Quality of Control (QoC)}. QoC of a TCPS is a relative measure of its quality compared to the quality of an ideal TCPS where an ideal TCPS is one with $1ms$ RTT, zero packet loss and zero jitter. Notice that $\Delta_{opt}$ for the ideal TCPS will be equal to its RTT, i.e., $1ms$. %This is in accordance with existing literature that expects an ideal TCPS system to have an RTT of $1ms$. 
Let $t_{r,ideal}$ $\triangleq$ $t_{r,ideal}(1ms)$ be the rise time of the corresponding $\Delta_{opt}$ curve. We then define QoC for the TCPS under test as,
%
%
%We propose a metric to evaluate TCPS. We call this metric, an indicator of TCPS control loop quality, \textit{Quality of Control (QoC)}.
%
%QoC is computed by comparing the TCPS under test with an ideal TCPS implementation. 
%
%We use the following equation for this; we use $t_r(\Delta)$ to explicitly show dependence of $t_r$ on $\Delta$. 

\begin{equation} 
QoC = \log_{10} \left(\dfrac{t_{r,ideal}}{t_r( \Delta_{opt})} \right)
\label{eqn_tcpsQEM}
\end{equation}
Where, $t_r(\Delta_{opt})$ is the rise time of the $\Delta_{opt}$-curve of the TCPS under test. We use log$_{10}$ for the convenience of representing $t_r$ ratio. We expect rise time ratio in \eqref{eqn_tcpsQEM} to assume a wide range as the latencies associated with TCPS components~(e.g., network latency) can vary widely. This is why we use a logarithmic scale to specify QoC.    

We now describe how we compute $t_{r,ideal}$. Observe that $t_{r,ideal}$ depends on $k_p$ and $k_1$ through their product. We set $k_p k_1 = 1$ as suggested in Section~\ref{sec_designOfEvalutionMetric}. From Figure~\ref{fig_ControlParameterSettingsZoomIn}, we see that, for an ideal TCPS with $k_pk_1 = 1$, the PI controller takes three loop times~($\approx 3\Delta_{opt} = 3ms$)  from $x = 50cm$ to $x = 53cm$ to correct the step change. Also $t_{r,ideal}$ is $1.5$ loop times~$\approx 1.5ms$.

Observe that QoC of the TCPS under test will be positive if it has $t_r(\Delta_{opt}) <  1.5 ms$ and negative if it has $t_r(\Delta_{opt}) > 1.5 ms$. The metric intuitively indicates how fast (or slow) the operator can control the teleoperator using the haptic feedback without introducing significant control glitches at the teleoperator side.

\paragraph*{Discussion}
We have found several notions of quality of control, also referred to as QoC,  in  real-time control literature \cite{ref_Buttazzo2007, ref_Buttazzo2004, ref_Tian2011, ref_Bini2008, ref_Aminifar2016, ref_Aminifar2018, ref_Cervin2002}. These definitions of QoC refer to objective metrics used for measuring the performance of generic control systems. QoC in the present work is, however, different from these existing notions in terms of its objective, usefulness and measurement method. We describe these differences in Section~\ref{sec_QoCInLiterature}. 

\subsection{Methodology to Determine QoC}
\label{sec_qocDetermination}

The above definition of QoC is on the presumption that each TCPS~(for a given $k_p$, $k_1$) has an optimal loop wait time, $\Delta_{opt}$. Also, $\Delta_{opt}$, which is a function of TCPS RTT, packet drop rate, etc., can be estimated through a sequence of step response experiments. However, for any TCPS, its RTT, packet drops, etc., may vary with time. Hence, QoC, if measured as defined above, will vary with time. One can define QoC for a TCPS for the worst-case RTT and packet drops, but such a measure will present a pessimistic picture of the system and will be of little consequence in practice. In most of the applications, we want good responses for a prescribed fraction, say $g_{spec}$, of time. We call $g_{spec}$ the desired goodness percentage. For instance, $g_{spec}$ could be 0.99 for critical applications and smaller for others. Here, we give an alternate characterization of the quality of control as a function of the desired goodness percentage; we call it $\overline{QoC}(g_{spec})$, We also describe a way to measure it in practice.   

Recall that $\Delta_{opt}$ is the minimum value of $\Delta$ that yields a good response. We define $\overline{\Delta_{opt}}(g_{spec})$ to be the minimum $\Delta$ that yields a good response curve $g_{spec}$ fraction of time. Note that $\overline{\Delta_{opt}}(g_{spec})$ will be large for a large $g_{spec}$. We can obtain  $\overline{\Delta_{opt}}(g_{spec})$ as follows. We choose a small $\Delta$ and run the step response experiment a large number of times, say $m$ times. We determine the fraction of times, say $g \approx (g > 0)$, good response curves are obtained. The parameter $m$ is selected to limit the $95\%$ confidence interval range of $g$ to be within $\pm5\%$. We now increase or decrease $\Delta$ depending on whether $g$ is smaller or larger than $g_{spec}$ and repeat the above experiments until $g = g_{spec}$. The final value of $\Delta$ is the desired $\overline{\Delta_{opt}}(g_{spec})$. We define $t_r(\overline{\Delta_{opt}}(g_{spec}))$ to be the average value of $t_r$ over all the good response curves corresponding to $\overline{\Delta_{opt}}(g_{spec})$. Finally, we define,

\begin{equation} 
\overline{QoC}(g_{spec}) =   \log_{10} \left(\dfrac{t_{r,ideal}}{t_r(\overline{\Delta_{opt}}(g_{spec}))} \right)
\label{eqn_QoCWithPspec}
\end{equation}

\subsection{Factoring in the Effect of Data Packet Size}

\begin{figure}[!htbp]
\centering
\includegraphics [width=0.9\linewidth]{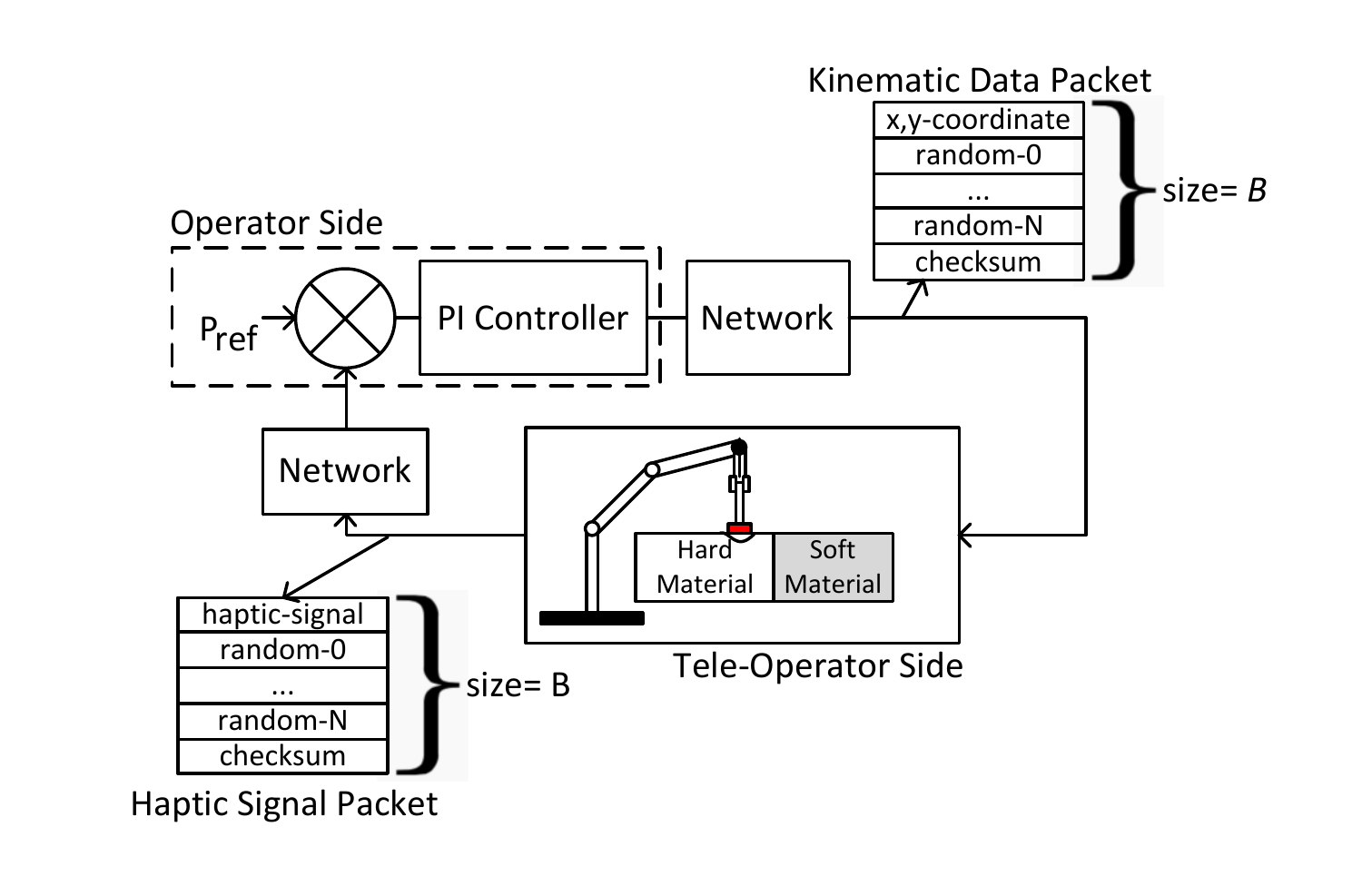}
\caption{Evaluation framework to account for variable data packet size.}
\label{fig_controlModelwithEnhancedPacketSize}
\end{figure}

In the evaluation methodology discussed we use an approximate model for an end-to-end TCPS. We assume there are only one haptic sensor and two degree of freedom for the robot. Here the forward kinematic and backward haptic signal packets are thus limited in size. 
However, a real TCPS will have a robot with multiple degrees of freedom and an array of haptic sensors which increase the data packet size. 
The increase in the data packet size can affect the end-to-end system latency, jitter, and packet errors. We thus believe for evaluation, we have to account for the real data packet size. For this, we propose a modified evaluation framework (see~Figure~\ref{fig_controlModelwithEnhancedPacketSize}). We modify the code snippets at the operator and tele-operator side to simulate a defined packet of size $B$ bytes. We do this by appending the haptic and kinematic data packets with random bytes and the corresponding checksum. At the operator and tele-operator side, we accept the packets only upon a matching checksum. This enables us to capture the effects of packet errors in the forward and backward TCPS channels. For evaluating QoC, we set $B$ to match the data packet size of a real-world TCPS. 

\section{QoC vs. Operator Kinematics}
\label{sec_VmaxSection}
%\todo[inline]{Question: Should we define $Vmax(g_{spec})$ instead of $V_{max}$}
For a human operator, the maximum allowed hand speed, $V_{\max}$, to avoid cybersickness is limited by two factors: the natural limit on human hand speed ($1m/s$) and the quality of the TCPS \textit{kinematic-video} loop.   The latter is not the limiting factor in an ideal TCPS. More specifically, an  ideal TCPS with $RTT = 1ms$ and $\overline{QoC}(1) = 0$ supports $V_{max} = 1m/s$.\footnote{For an ideal TCPS with no randomness, for all $g$, $\overline{QoC}(g) = QoC = 0$.} A TCPS with lower $\overline{QoC}(g_{spec})$~(for the given $g_{spec}$) will support smaller hand speeds only, i.e., it will have $V_{max} < 1m/s$.  In this section, we first derive the  relation between $\overline{QoC}(g_{spec})$ and $V_{max}$. We then validate this relation through simulation.

%For a human operator, the maximum allowed hand speed, $V_{\max}$, is limited by two factors: the natural limit on human hand speed ($1m/s$) and the quality of the TCPS. 
%For an ideal TCPS with $1ms$ RTT ($QoC = 0 \text{dB}_{ideal}$), $V_{\max}$ takes the value of $1m/s$, the natural limit on human hand speed \cite{6755599}. 
%
%For TCPS with lower performance ($QoC < 0 \text{dB}_{ideal}$), $V_{\max}$ has to be less than $1m/s$. This is to avoid the human operator from experiencing cybersickness \cite{6755599}. In this section, we first derive the relation to estimate $V_{\max}$ from  QoC. We then validate the relation using simulations.

\subsubsection{Maximum hand speed}

\begin{figure}[!htbp]
\centering
\includegraphics [width=0.85\linewidth]{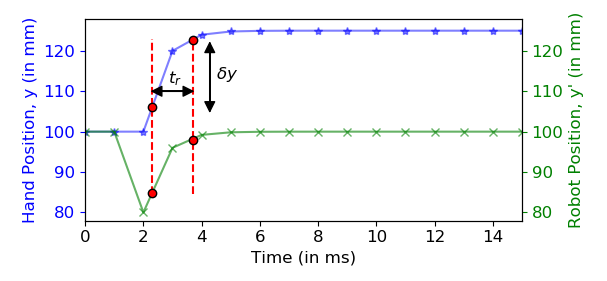}
\caption{Step response plots of the operator's hand position, $y$, and haptic signal for an ideal TCPS with zero packet drop and $RTT_{max}=1ms$, and for a data sampling interval of $1ms$.}
\label{fig_VTr}
\end{figure}

For an ideal TCPS with $RTT=1ms$ and no packet drop and jitter, the step response graph of robot $y$-coordinate i.e., plot of $y'$, and the graph of operator's hand position i.e., plot of $y$,  controlling the $y$-coordinate of the robotic arm, sampled at every $1ms$, will have the same rise time $t_r$; see Figure~\ref{fig_VTr}. For the plot, we choose $Y_{ref}=100mm$ and use $\delta y$ to represent the change in operator's hand position.  This condition (the rise time being the same) prevails in generic TCPS also, provided we use $\Delta \geq \overline{\Delta_{opt}}(g_{spec})$ . In the case of a generic TCPS, $t_{r}$ and $\delta y$ are replaced with ${t_r( \overline{\Delta_{opt}}(g_{spec})}$ and $\delta y(\overline{\Delta_{opt}}(g_{spec}))$, respectively.

We define the maximum allowed operator's hand speed $V_{\max}$ as $dy/dt$. For generic TCPS, 
%We define the maximum speed of the operator's hand $V_{\max}$ as $dy/dt$. For generic TCPS, $V_{\max}$ is given by  
%For a generic TCPS, the maximum speed of the operator's hand, $V_{\max}$, is given by 
%
\begin{equation} 
V_{\max} = \dfrac{dy}{dt} = \dfrac{\delta y(\overline{\Delta_{opt}}(g_{spec}))}{{t_r(\overline{\Delta_{opt}}(g_{spec}))}} 
\label{eqn_V_1}.
\end{equation}
Substituting $t_r(\Delta_{opt})$ from \eqref{eqn_tcpsQEM} in \eqref{eqn_V_1}, we get
\begin{equation} 
V_{\max} = \dfrac{\delta y(\overline{\Delta_{opt}}(g_{spec}))}{t_{r,ideal}} \times 10^{\overline{QoC}(g_{spec)}}.
\label{eqn_V_2}
\end{equation}

\begin{figure}[!htbp]
\centering
\includegraphics [width=0.85\linewidth]{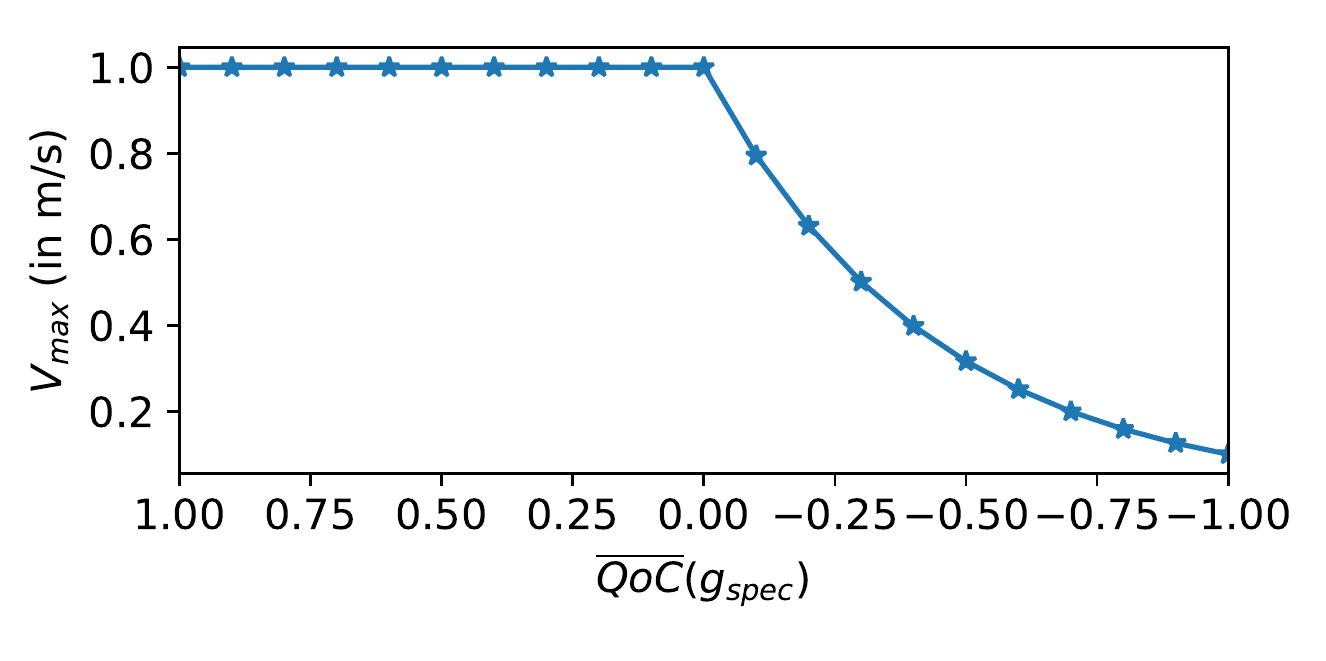}
\caption{Graph relating $V_{\max}$ and QoC.}
\label{fig_VGraph}
\end{figure}

Applying \eqref{eqn_V_2} to the ideal TCPS, i.e., substituting $V_{\max} = 1m/s$ and $\overline{QoC}(g_{spec})=0$, we obtain $\delta y(\overline{\Delta_{opt}}(g_{spec}))/ t_{r,ideal}$ = $1m/s$, a constant. Using this and also noting that $V_{max}$ is limited  to $1m/s$, the natural limit on human hand speed, we  obtain the following simple relation between $V_{max}$ and $\overline{QoC}(g_{spec})$ in any TCPS . We illustrate this relation in Figure~\ref{fig_VGraph}.

\begin{equation} 
%V_{\max} = \max \{1, \times 10^{QoC}\}.
V_{\max} = \min \{1, 1 \times 10^{\overline{QoC}(g_{spec})}\} m/s.
\label{eqn_speedOfHandKinematics}
\end{equation}
\subsubsection{Simulation Results}
\begin{figure}[!htbp]
\centering

\begin{subfigure}[b]{1\linewidth}
\centering
\includegraphics [width=0.85\linewidth]{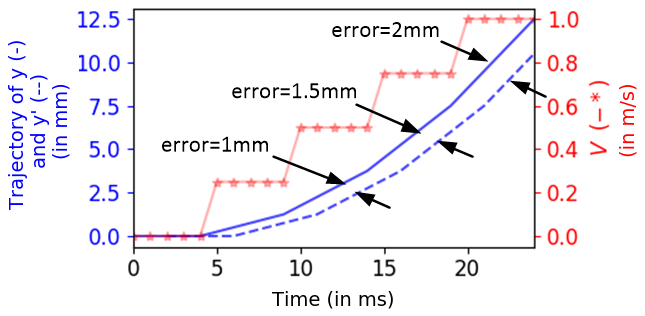}
\caption{For a TCPS with $\overline{QoC}(1)=-0.3$. Here error exceeds $1mm$ for $V > (V_{\max}=0.5m/s)$.}
\label{fig_VEvaluation}
\end{subfigure}

\begin{subfigure}[b]{1\linewidth}
\centering
\includegraphics [width=0.85\linewidth]{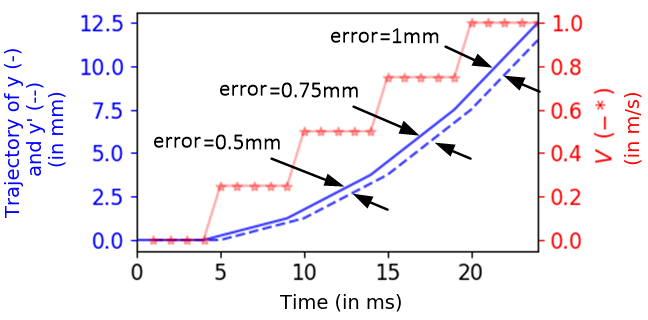}
\caption{For a TCPS with $\overline{QoC}(1)=0$. Here error remains within $1mm$ upto $V = (V_{\max}=1m/s)$.}
\label{fig_VEvaluationCase2}
\end{subfigure}

\caption{Plots describing how the error between the operator's hand position, $y$, and robot position, $y'$, vary with operator hand speed, $V$, for TCPS with different QoCs.}

\end{figure}

In any TCPS applications, the human operator is expected to restrict his hand speed to $V_{\max}$ a limit posed by the underlying TCPS. If the hand speed exceeds, the operator may experience cybersickness due to visible error ($> 1mm$) between his/her hand position and the robot position seen in the video feedback (see Figure~\ref{fig_tcpsRttEffect}). We demonstrate this in Figure~\ref{fig_VEvaluation} and \ref{fig_VEvaluationCase2}. Here, $y$ represents the operator's hand position and $y'$ represents the robot position displayed on the operator side screen. For Figure~\ref{fig_VEvaluation}, we consider a TCPS with $\overline{QoC}(1)$ which corresponds to $V_{\max} = 0.5m/s$. As long as the operator maintains his hand speed, $V$, within $0.5m/s$, the error between $y$ and $y'$ is bounded within $1mm$. This error crosses $1mm$ when $V$ exceeds $0.5m/s$. For Figure~\ref{fig_VEvaluationCase2}, we consider a TCPS with $\overline{QoC}(1) = 0$ which corresponds to $V_{\max} = 1m/s$. Here the error between $y$ and $y'$ remains within $1mm$ up to $V=1m/s$. 

\begin{comment}
\begin{figure}[!htbp]
\centering
\includegraphics [width=0.85\linewidth]{./Attachments/chapEvaluationOnly/fig_ShkEvaluation.png}
\caption{For a TCPS with $\overline{QoC}(1)=-0.3$, error between the operator's hand position, $y$, and robot position, $y'$, exceeds $1mm$ when $V$ crosses $(V_{\max}=0.5m/s)$.}
\label{fig_VEvaluation}
\end{figure}
\end{comment}

\section{QoC in Literature: Definitions and Differences}
\label{sec_QoCInLiterature}
We find that QoC is a term often used in real-time control literature. It refers to objective metrics used for measuring the performance of control systems. Several definitions of $QoCs$ are found in literature \cite{ref_Buttazzo2007, ref_Buttazzo2004, ref_Tian2011, ref_Bini2008, ref_Aminifar2016, ref_Aminifar2018, ref_Cervin2002}. Broadly, we classify these definitions into two groups depending on the performance measures authors use to determine QoC.

\begin{figure}[!htbp]
\centering
\includegraphics [width=1\linewidth]{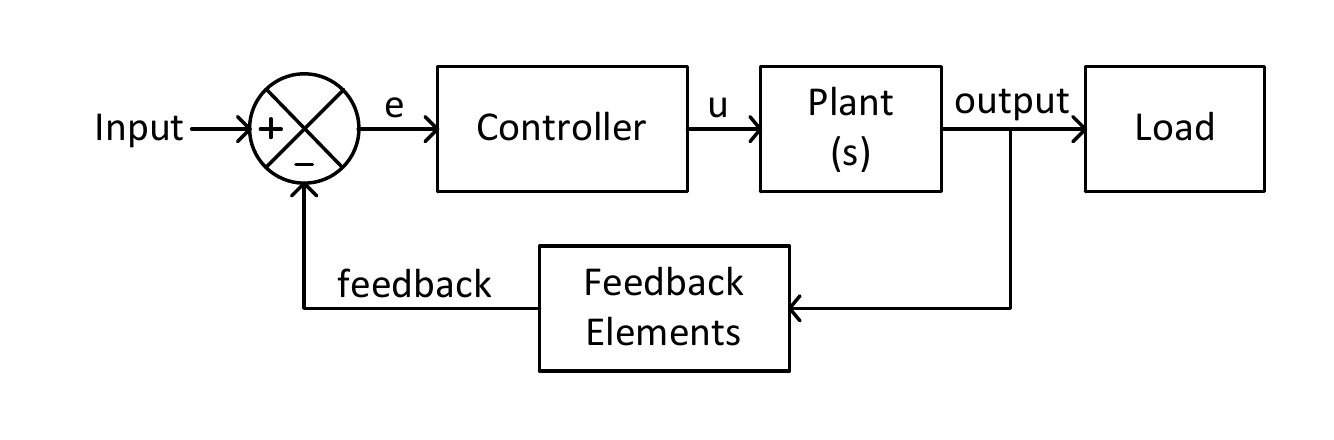}
\caption{Block representation of a generic control system.}
\label{fig_ControlSystemBlock}
\end{figure}

\paragraph*{Definition-1} To determine QoC, authors measure the integral of error, $e$, after simulating a step disturbance in the control loop shown in Figure~\ref{fig_ControlSystemBlock}. Authors may use different approaches to measure integral of error like integral of absolute error (IAE), integral of square error or integral of weighted absolute error. For example, authors in \cite{ref_Buttazzo2007, ref_Buttazzo2004} measure IAE to determine QoC as follows. % (see~\eqref{eqn_IAE}). 

\begin{subequations} 
\begin{align}
IAE = \int_{0}^{\infty}|e(t)|dt \\
QoC = \frac{1}{IAE}
\end{align}
\label{eqn_IAE}.
\end{subequations}

\paragraph*{Definition-2} To determine QoC, authors measure the quadratic cost, $J$, instead of integral of error \cite{ref_Tian2011, ref_Bini2008, ref_Aminifar2016, ref_Aminifar2018, ref_Cervin2002}. Equation\eqref{eqn_quadraticCost} represent the generic form of $J$. $u$ is the input to the plant, $s$ is the plant state variable, and $R$ and $Q$ are weighing inputs. QoC evaluation using $J$ has the advantage that it accounts for both the error and the energy consumed by the controller.

\begin{equation} 
J = \frac{1}{2}\int_{0}^{\infty}Ru(t)^2+Qs(t)^2dt
\label{eqn_quadraticCost}.
\end{equation}

In Table~\ref{tbl_QoCLiterature}, we list the essential difference between our definition of QoC and QoC definitions in the literature. 

%In the following section, we list the essential difference between our definition of QoC and QoC's in the existing literature. 

% Please add the following required packages to your document preamble:
% \usepackage{booktabs}
% Please add the following required packages to your document preamble:
% \usepackage{booktabs}
\begin{table*}[]
\centering
\caption{Essential differences between our definition of QoC and QoC definitions in real-time control literature.}
\label{tbl_QoCLiterature}
\begin{tabular}{@{}l|c|c@{}}
\toprule
\multicolumn{1}{c|}{}         & \textbf{QoC (Our Definition)}                                                                                    & \textbf{QoC (in Literature)}                                                                                          \\ \midrule
\textbf{Objective}             & \begin{tabular}[c]{@{}c@{}}Evaluates plant (TCPS) performance \\ for a given controller and load.\end{tabular} & \begin{tabular}[c]{@{}c@{}}Evaluates controller performance\\ for a given plant and load.\end{tabular}               \\ \midrule
\textbf{Performance Measure}    & \begin{tabular}[c]{@{}c@{}}Rise time of  good \\ step response curves.\end{tabular}                             & \begin{tabular}[c]{@{}c@{}}Time domain integration of \\ step response signals.\end{tabular} \\ \midrule
\textbf{Physical Significance} & Useful in determining $V_{max}$.                                                                                 & No physical significance.                                                                                             \\ \bottomrule
\end{tabular}
\end{table*}
\begin{itemize}
\item  Although QoCs in the literature and QoC presented in our work both measure control performance, their objectives are different. In literature, authors use QoC to assess the quality of the controller for a given plant and a given load. However, we use QoC with the intent of assessing the quality of the plant, i.e. the TCPS. It is for this reason, we fix the controller and the load by fixing $k_p$, $k_1$ and $k_2$. 

\item In the literature, authors, determine QoC by time-domain integration of signals (e.g. $IAE$, $J$).  QoC's in the literature thus cannot comprehend the real dynamics of a control system \cite{ref_Buttazzo2007}. For instance, it is possible that two different control systems, one that generates a  spiky transient response with a low steady-state error and the other that generates a non-spiky transient response with a high steady-state error, can both yield the same $IAE$. In generic control systems, this might be acceptable but not for TCPS applications, where the presence of spikes can hamper the remote operation. For this reason, we use $t_r(\overline{\Delta_{opt}}(g_{spec}))$, the rise time of the tuned step response curves as the performance measure to determine QoC. Recall that tuning of step response curves ensures $\Delta_{opt}$ and thus $t_r$ to account for both the transient and steady-state components of the error.

\item The performance measures the authors use to determine QoCs in the literature devoids the resultant metric from having a physical significance, i.e., for a TCPS, they cannot comment on the recommended maximum operator hand speed, $V_{max}$. We determine QoC by measuring the rise time of the tuned good step response curves and describe in  Section~\ref{sec_VmaxSection} how to relate $V_{max}$ and QoC. 

\end{itemize}

%Although QoC's in literature and the QoC presented in our work both measure control performance, their objectives are different. QoC in literature measure control performance to assess the quality of the controller for a given plant and given load. However, in our work, we use QoC with the intent of measuring the quality of the plant (i.e., the TCPS). For this reason, we fix the controller and load by fixing $k_p$, $k_1$ and $k_2$. 
%
%Performance measures used for existing QoC’s are based on integration of error profiles. The existing QoC’s thus cannot comprehend the real dynamics of a control system i.e, they cannot be used to differentiate a spiky response with low steady state error and a non spiky response with a high steady state error. In generic control systems this might be acceptable but not for TCPS applications like telesurgery. (Reason why for determining QoC, we demand tuning DELTA for good step curves)
%
%Performance measures used for existing QoC’s devoid QoC from having a physical significance, i.e, for TCPS, they cannot say what should be the recommended maximum operator hand speed. (Reason why for determining QoC, we use rise time of step response curve as a performance measure)

\section{QoC Performance Curve}
\label{sec_qocPerformanceCurve}
\begin{figure}[!htbp]
\centering
\includegraphics [width=0.85\linewidth]{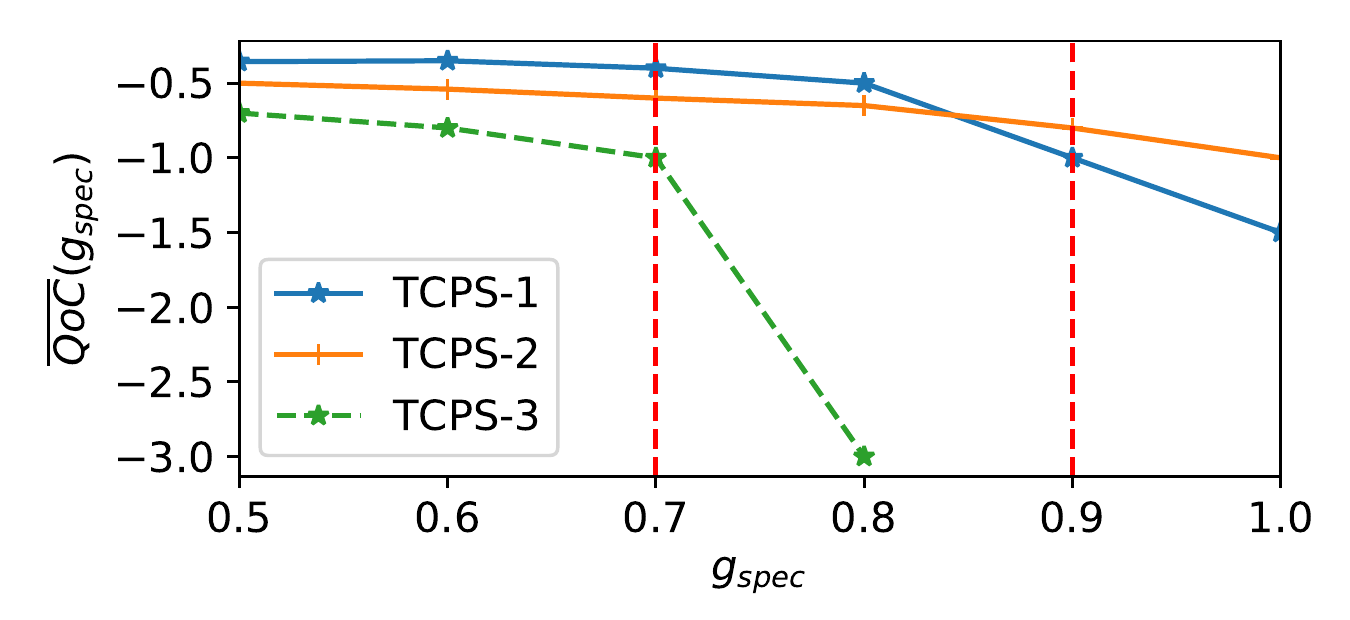}
\caption{Sample QoC performance curves for three different TCPS.}
\label{fig_qocPerformanceCurve}
\end{figure}

When comparing performance of different TCPS using QoC, we use a common $g_{spec}$ (i.e., $\overline{QoC}(g_{spec})$. Here $g_{spec}$ is specified by the TCPS application. However, when vendors publish QoC for TCPS targeting multiple applications, listing QoC for one $g_{spec}$ is not enough. As a solution, we propose publishing the QoC performance curve for TCPS. The curve illustrates how QoC varies with $g_{spec}$. In Figure~\ref{fig_qocPerformanceCurve}, we plot sample QoC performance curves for three different TCPS. From the curves, we conclude the following; (i) for applications that demand $g_{spec} =0.7$, TCPS-1 has better performance compared to TCPS-2 and TCPS-3 (ii) for applications that demand $g_{spec}=0.9$, TCPS-2 has the better performance compared to TCPS-1. For this application, we cannot consider TCPS-3 as its QoC for $g_{spec}>0.8$ is not specified.

\section{Performance Evaluation}
\label{sec_evaluation}
In this section, we demonstrate how to use QoC to evaluate the overhead of a TCPS testbed, evaluate the quality of a TCPS network under different traffic conditions and to validate the effectiveness of QoC in quantifying cybersickness. 

\subsection{Evaluating the Testbed Overhead}

We determine the overhead introduced by the testbed in Figure~\ref{fig_testBed} in terms of QoC. We find the overhead of the testbed framework without the network emulator to be $\overline{QoC}(0.9)= -0.35$. With network emulator, the overhead increases to $\overline{QoC}(0.9) = -0.92$.

\begin{figure}[!htbp]
\centering
\includegraphics [width=0.8\linewidth]{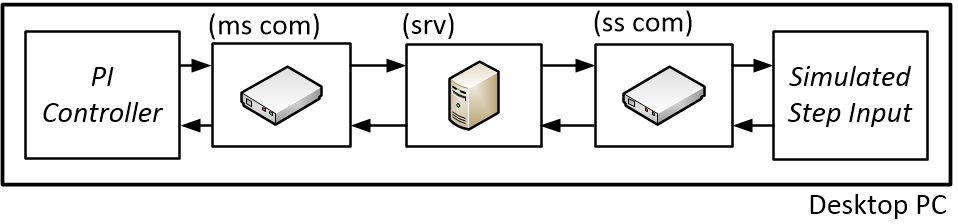}
\caption{Setup for evaluating the testbed overhead. Arrows mark the forward kinematic and backward haptic signal flows}
\label{fig_testbedOverheadSetup}
\end{figure}

\paragraph {Overhead of Testbed Framework}
To evaluate the overhead introduced by the testbed framework, we realize the components \textit{ms com}, \textit{srv} and \textit{ss com} in a single desktop PC (Processor:Intel-Core-i5, Core Count: 4, Processor Frequency: 3.4Ghz, Memory: 3.7GiB) with $<10\%$ and $<20\%$ CPU and RAM utilizations, respectively (see Figure~\ref{fig_testbedOverheadSetup}).
To ensure the results are representative of the testbed overheads alone, (i) we retain minimal code in the testbed components to enable only the needed inter-component communication and (ii) we substitute the teleoperator side using a code snippet that also simulates the step input (to avoid accounting for teleoperator side component overheads). Following the steps (i) and (ii) results in the evaluation frameworks defined for a haptic setting and for a non-haptic setting to appear indistinguishable. i.e., irrespective of which evaluation framework we use, the end QoC will be the same.

\begin{figure}[!htbp]
\centering

\begin{subfigure}[b]{1\linewidth}
\centering
\includegraphics [width=0.81\linewidth]{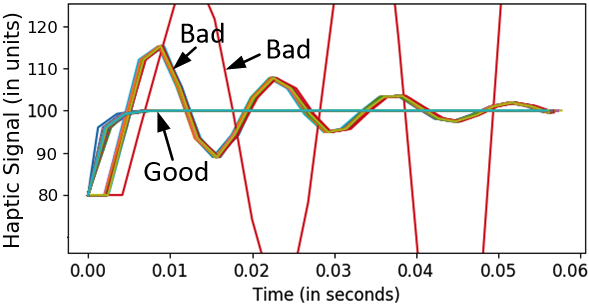}
\caption{Plot of 1000 step curves for $\Delta = 1.9ms$. Here $90.6$\% of step curves are good.}
%\label{fig_stepCurvesExperimentFail}
\end{subfigure}

\begin{subfigure}[b]{1\linewidth}
\centering
\includegraphics [width=0.81\linewidth]{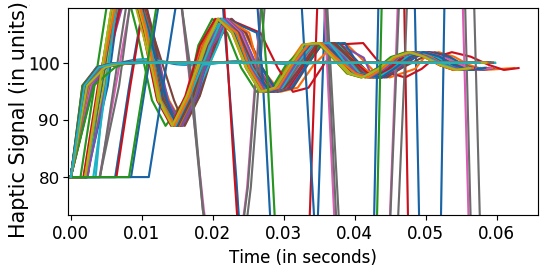}
\caption{Plot of 1000 step curves for $\Delta=1.8ms$. Here $80$\% of step curves are good..}
%\label{fig_stepCurvesExperimentFail}
\end{subfigure}
\caption{Plot of 1000 step curves for two different $\Delta$. We find, ${g>0.9}$ only when ${\Delta \geq 1.9ms}$. Thus we determine ${\overline{\Delta_{opt}}(0.9)=1.9ms}$ for a $g_{spec}=0.9$.}
\label{fig_stepCurvesExperiment}
\end{figure}

We run the experiment to extract the step response curves for different values of $\Delta$. For each $\Delta$, we extract $m=1000$ step curves.  (see Figure~\ref{fig_stepCurvesExperiment}). 

\paragraph*{Result} 
We find $\overline{\Delta_{opt}}(0.9) = 1.9ms$,  $t_r(\overline{\Delta_{opt}}(0.9))=3.364ms$ ($\pm 0.21\%$) and $\overline{QoC}(0.9) = -0.35$. The metric indicates that the testbed implemented on the selected desktop PC and OS configuration is inadequate to realize ideal TCPS with $1ms$ RTT. We either have to optimize the socket I/O calls or optimize background processes or replace the hardware. Note that, $QoC<0$ does not mean that the TCPS implementation is not useful.  It means the operator has to restrict his/her maximum hand speed to $V_{\max}$ during teleoperation. From \eqref{eqn_speedOfHandKinematics}, $V_{\max} = 0.44m/s$.

%\underline{Result:} We find $\Delta_{opt} = 2ms$, mean $t_r \mid _{\Delta_{opt}}=3.364ms$(+/-$0.21\%$) and $p=90.6\%$(+/-$1.645\%$). This corresponds to $QoC = \left. -0.35  dB_{ideal} \; \right \vert _{p = 90\%}$.

\begin{comment}
\begin{table}[!htbp]
\centering
\caption{Overhead of Testbed Framework (Experimental Result)}
\begin{tabularx}{\linewidth}{ccc}

\toprule
$\Delta_{opt}$ & $t_r \mid _{\Delta_{opt}}$  &  QoC \\
\midrule

$2ms$  & $3.364ms$+/-$0.21\%$ & $\left. -0.35  dB_{ideal} \; \right \vert _{p = 90\%+/-1.645\%}$ \\
          
\bottomrule
\end{tabularx}
\label{tbl_resultTestbedFramework}
\end{table}
\end{comment}

\begin{figure}[!htbp]
\centering
\includegraphics [width=1\linewidth]{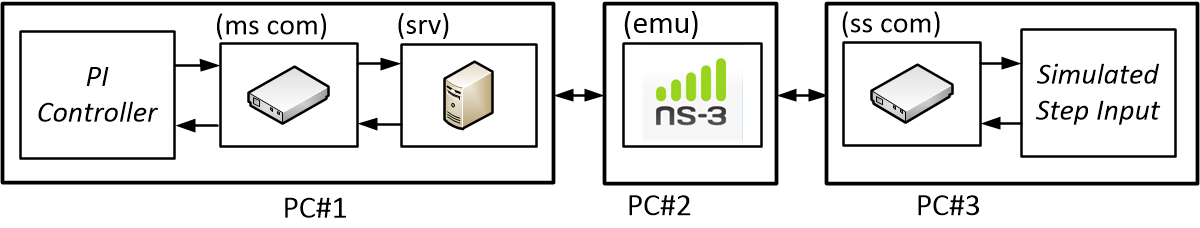}
\caption{Setup for evaluating the testbed overhead with network emulator, \textit{emu}.}
\label{fig_testbedNs3OverheadSetup}
\end{figure}

\paragraph{Overhead of Integrating Network Emulator }
To understand the overall overhead of the testbed with the network emulator in place, we modify the setup in Figure~\ref{fig_testbedOverheadSetup}. We evaluate QoC by placing the components \textit{ms com} and \textit{srv} in PC\#1, \textit{emu} running the ns-3 code in PC\#2 and \textit{ss com} in PC\#3 (see~Figure~\ref{fig_testbedNs3OverheadSetup}). The PCs are connected using point-to-point gigabit Ethernet links to interconnect the components, as shown in Figure~\ref{fig_testBed}. In ns-3, we emulate an ideal point-to-point link of zero latency, and zero packet drops to ensure the QoC measurement captures the testbed overhead alone and not the effects of any network components in ns-3.

\paragraph*{Result} We find $\overline{QoC}(0.9) = -0.92$ indicating that incorporating the emulator block and running testbed components in different PCs increases the testbed overhead. From \eqref{eqn_speedOfHandKinematics}, this restricts $V_{max}$ to $0.12m/s$.

\subsection{QoC Assessment for Wifi Links}

\begin{figure}[!htbp]
\centering
\includegraphics [width=0.8\linewidth]{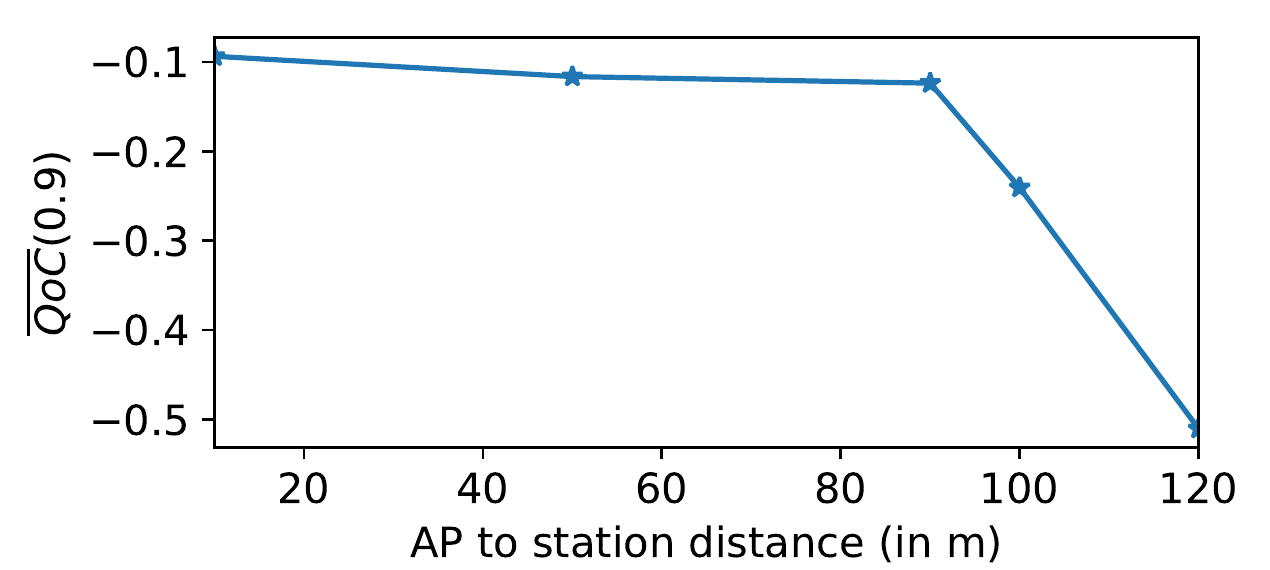}
\caption{QoC values for the IEEE802.11n link for different Access Point (AP) to station distances.}
\label{fig_wifiChar}
\end{figure}

We assess the suitability of IEEE 802.11n (5GHz band) as the last-mile radio link for TCPS using the setup in Figure~\ref{fig_testbedNs3OverheadSetup}. Here, the IEEE 802.11n link is simulated in ns3. We determine the QoC value of the link by correcting for the testbed overhead. We plot the QoC values in Figure~\ref{fig_wifiChar}. For implementing an ideal TCPS, the last-mile radio link is expected to have $RTT < 200us$ \cite{Fettweis:2014aa}. This transforms to a QoC requirement $>0.69  \text{dB}_{ideal}$. From Figure~\ref{fig_wifiChar}, we find the QoC of the radio link to be negative, indicating that its round trip latency is $>1ms$. Thus we may conclude that IEEE802.11n has a limited performance to realize radio links for TCPS that require almost ideal performance, i.e., to support $V_{\max}=1m/s$.  
\subsection{Evaluating the Impact of an Intercontinental Link}

We understand that intercontinental links are not suited for TCPS that require $V_{max}=1m/s$. However, we assess a real-world intercontinental link for TCPS to learn the $V_{max}$ it can support. For the experiments, we replace PC\#2 in Figure~\ref{fig_testbedNs3OverheadSetup} with an intercontinental link. We place PC\#1 and PC\#3 at two university locations in Asia and Europe. We find $\overline{QoC}(.5)= -2.99$ and $V_{\max}=0.001m/s$. The resulting low $V_{\max}$ at low $g_{spec}$ eliminates any practical use case for the implemented TCPS.

%this restrict the operator hand movement to $0.001m/s$. 
%The resultant QoC yields a very low $V_{max}= 0.001m/s$.  We thus believe that the intercontinental links in its current form are not suited for implementing TCPS. 

% \begin{figure}[!htbp]
% \centering
% \includegraphics [width=1\linewidth]{./Attachments/chapEvaluationOnly/fig_testbedIntercontinental.png}
% \caption{Experimental setup for evaluating the impact of real world intercontinental link on TCPS applications. Arrows mark the forward kinematic and backward haptic signal flows. }
% \label{fig_testbedIntercontinental}
% \end{figure}
%\todo[inline]{... pending}

\subsection{Comparing TCPS networks using QoC}
\begin{figure}[!htbp]
\centering
\includegraphics [width=0.8\linewidth]{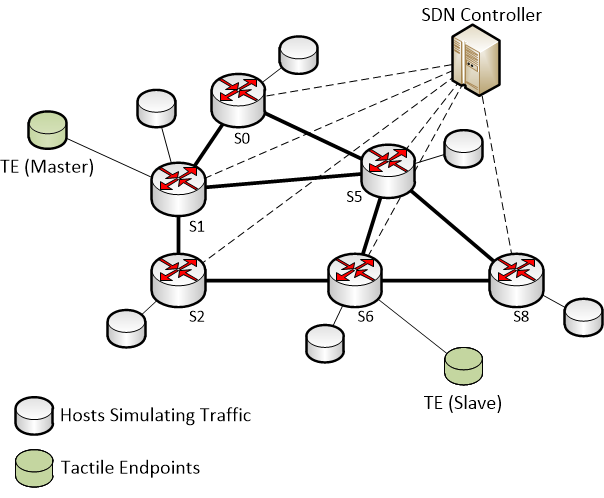}
\caption{Network topology simulated in Mininet.}
\label{fig_mininetNetwork}
\end{figure}

We demonstrate how to use QoC to compare networks with different traffic conditions and different tactile endpoint locations. For evaluation, the network  Figure~\ref{fig_mininetNetwork} is simulated using Mininet \cite{ref_mininet}. The network represents the north-west subset of the USNET-24 topology. It consists of six switches managed by a central SDN controller. Tactile endpoints TE(Master) and TE(slave) connects to switches S1 and S6. Each switch in the network connects to a host, and every host communicates with every other host bidirectionally to simulate external traffic. The communication data is generated using the Linux tool iPerf. The SDN controller determines the routes between the hosts. Spanning tree protocol is used to avoid looping issues. For performing the QoC experiment, TE(Master) runs the PI controller, and TE(Slave) simulates step change. All host to switch links are made ideal, i.e., the links are configured to be of very low latency and very high bandwidth. Latencies and bandwidth of all links interconnecting the switches are set as variables for the user to set. The Mininet simulation is run on a server to reduce overheads.

\subsubsection{Effect of External Traffic}

\begin{figure}[!htbp]
\centering
\includegraphics [width=0.8\linewidth]{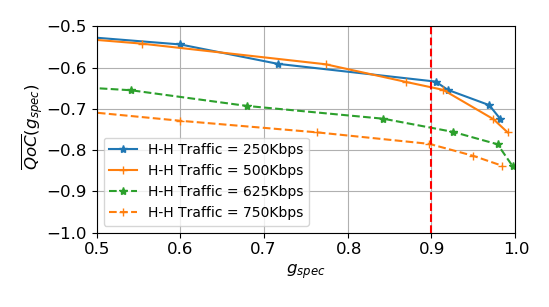}
\caption{QoC performance curves for different traffic conditions.}
\label{fig_mininetTrafficQoC}
\end{figure}

\begin{figure}[!htbp]
\centering
\includegraphics [width=0.8\linewidth]{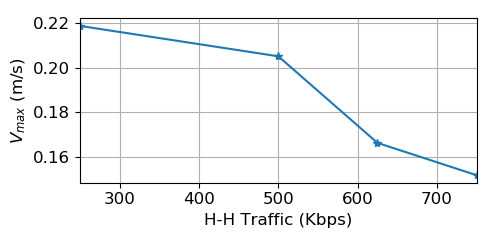}
\caption{$V_{max}$ for $g_{spec}=0.9$ for different traffic conditions.}
\label{fig_mininetTrafficVmax}
\end{figure}

In this experiment, for different traffic bandwidths simulated between host pairs, QoC is measured across the tactile endpoints.  We fix delay and bandwidth for the links interconnecting the switches to $0.1ms$ and $10Mbps$ respectively. Figure~\ref{fig_mininetTrafficQoC} shows the QoC performance curve for different simulated traffic bandwidths. The host to host traffic bandwidth (H-H Traffic) in the legend corresponds to the unidirectional bandwidth simulated between host pairs. Figure~\ref{fig_mininetTrafficVmax} shows the corresponding calculated $V_{max}$ for $g_{spec}=0.9$. 

\paragraph*{Result} 
(i) We find QoC curves corresponding to H-H Traffic of 250Kbps and 500Kbps to be close. Because for both these cases, the effective bitrate of the simulated traffic is not enough to throttle the bandwidth of the links connecting tactile endpoints. (ii) QoC performance curve has a higher slope at higher values of $g_{spec}$. Thus, aiming for higher values of $g_{spec}$ is costly w.r.t maintaining a higher value of QoC.

\subsubsection{Effect of Tactile Endpoint Locations}

\begin{figure}[!htbp]
\centering
\includegraphics [width=0.8\linewidth]{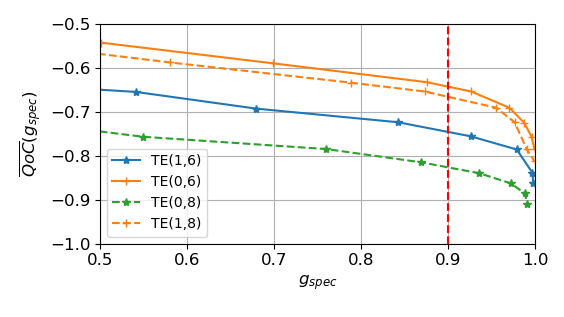}
\caption{QoC performance curves for different tactile end points.}
\label{fig_mininetEndPointsQoC}
\end{figure}

\begin{figure}[!htbp]
\centering
\includegraphics [width=0.8\linewidth]{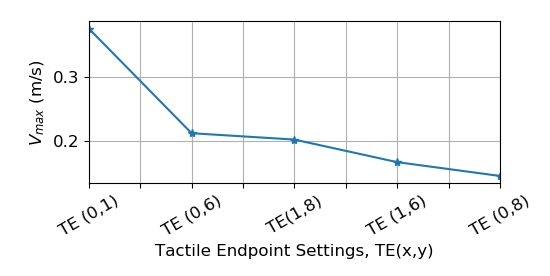}
\caption{$V_{max}$ for $g_{spec}=0.9$ for different traffic conditions.}
\label{fig_mininetEndPointsVmax}
\end{figure}

In this experiment, we connect the tactile endpoints, TE(Master) and TE(Slave) to different switch pairs, and measure QoC across the tactile endpoints. We fix a delay of 0.1ms and bandwidth of 10Mbps for the links interconnecting the switches. We simulate H-H Traffic=625Kbps between the host pairs. Figure~\ref{fig_mininetEndPointsQoC} shows the QoC performance curve for different tactile endpoint locations. TE(x,y) in the legend corresponds to the case where TE(Master) is connected to switch $x$ and TE(Slave) to switch $y$. Figure~\ref{fig_mininetEndPointsVmax} shows the corresponding calculated $V_{max}$ for $g_{spec}=0.9$. 

\paragraph*{Result} 
We find QoC performance curve corresponding to TE(0,8) has the lowest QoC values and hence, the lowest $V_{max}$. This is because the external traffic loads the route TE(Master)-S0-S5-S8-TE(Slave) more in comparison to other routes. The results are thus useful in predicting and how traffic in a network impact QoC performance depending on tactile endpoint locations.

\subsection{QoC Measurement and Validation} 
%\todo[inline]{edit pending ....}
In this experiment, we measure the combined QoC of a communication network and a connected robot in a non-haptic setting. We then use the QoC value to predict and quantify cybersickness.

\subsubsection{QoC Measurements}
\label{sec_QoCMeasurementwithVrep}
\begin{figure}[!htbp]
\centering
\includegraphics [width=0.68\linewidth]{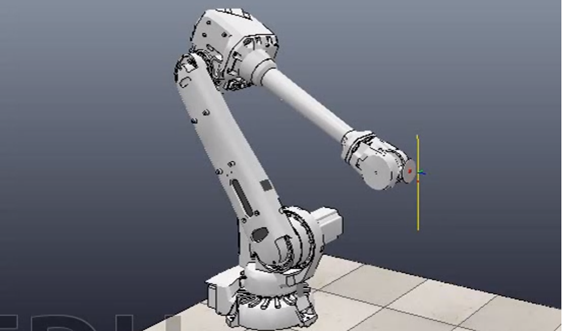}
\caption{Video frame captures the robot, IRB 4600, being actuated downwards along the vertical yellow line to simulate step change.}
\label{fig_vrep}
\end{figure}

%In this experiment, we measure the combined QoC of a communication network and a connected robot for a non-haptic setting. 
For evaluation, the network in Figure~\ref{fig_mininetNetwork} is simulated using Mininet \cite{ref_mininet}. TE(Master) runs the PI controller, and TE(Slave) simulates step change. The step change is simulated using the robot, IRB 4600, as described in Section~\ref{sec_evaluationFrameworkNonHapticSetting} (see Figure~\ref{fig_vrep}).  We use Virtual Robot Experimentation Platform (VREP) to simulate the robot \cite{ref_vrep}. We run VREP in TE(Slave) in the headless mode configuration, i.e., without GUI, to minimize the simulation overhead. A custom written python script interfaces the network socket of TE(Slave) with VREP. We configured the delay and bandwidth of the links interconnecting the switches to $5ms$ and $10Mbps$ respectively. H-H Traffic is disabled.

\paragraph*{Result}
We measure QoC with and without the robot in place. Without the robot, we find $\overline{QoC}(1) =  -1.42$, and with the robot, we find $\overline{QoC}(1) = -1.7$. From \eqref{eqn_speedOfHandKinematics}, this restricts $V_{max}$ to $0.02m/s$. %Results indicate a significant drop in performance with the addition of robot.

\subsubsection{QoC Validation}

If QoC of the TCPS and dynamics of the operator's hand movements are known a priori, we can predict and quantify the operator side cybersickness using the QoC-$V_{max}$ relation of \eqref{eqn_speedOfHandKinematics} as follows. For example, if QoC of a network is $\overline{QoC}(1) = -0.1$ (i.e, $V_{max}=0.1m/s$) and histogram of the operator's hand velocity suggest that the velocity is less than $0.1m/s$  for $90\%$ of the time, we predict $E$ to be $90\%$. Here $E$ is used to quantify cybersickness. We define $E$ as the percentage time during a TCPS operation, the error between the position of the operator's hand and the robot is within $|1mm|$.

\begin{figure}[!htbp]
\centering
\includegraphics [width=0.8\linewidth]{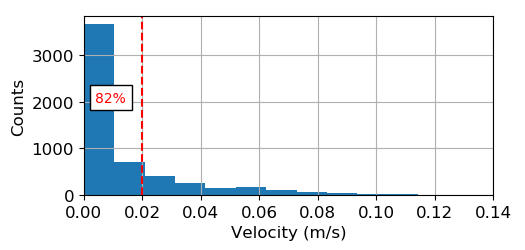}
\caption{Histogram of operator's hand velocity for $F_s=30Hz$. For 82\% of the time hand velocity is less than $V_{max} = 0.02m/s$. }
\label{fig_histDavinci}
\end{figure}

For validating the above claim, we use the TCPS in Section~\ref{sec_QoCMeasurementwithVrep}, whose QoC and $V_{max}$ is known ($\overline{QoC}(1) = -1.7$ and $V_{max} = 0.02m/s$). In the experimental setup, the operator side replays data corresponding to the hand movements of a surgeon performing the suturing operation \cite{ref_Gao2014}. At the teleoperator side, the robot (in VREP) replicates the surgeon's hand movement. Figure~\ref{fig_histDavinci} shows the histogram of the operator's hand movement for a sampling frequency of $F_s=30Hz$. We find that for $82\%$ of the time, the velocity is less than $V_{max}$ of $0.02m/s$. We thus expect $E$ to be $82\%$. In Table~\ref{tbl_qocValidation}, we list the expected and the measured values of $E$ for different $F_s$. We find the measured results to be very close to the expectation.

In the experiments, to measure $E$, at every $1/F_s$ seconds, the robot position is fed back to the operator side. At the operator side, at every instance of receiving the feedback, the error is calculated. For this, we subtract the robot position from the estimated position of the operator's hand at the receiving instance. From the error values, we find $E$ by determining the percentage of time the error values stay within $|1mm|$. For this, we plot the normalized PDF of the error values and find the area under the curve between $-1mm$ and $1mm$.

% Please add the following required packages to your document preamble:
% \usepackage{booktabs}
% Please add the following required packages to your document preamble:
% \usepackage{booktabs}
\begin{table}[!htbp]
\centering
\caption{Expected and measured values of $E$ for different values of $F_s$.}
\begin{tabular}{@{}c|c|c@{}}
\toprule
$F_s$ (Hz) & Expected $E$ (\%) & Measured $E$ (\%) \\ \midrule
40 & 77           & 81           \\ \midrule
30 & 82           & 86           \\ \midrule
20 & 88           & 92           \\ \bottomrule
\end{tabular}
\label{tbl_qocValidation}
\end{table}

\section{Conclusion}
\label{sec_conclusion}

We have proposed a comprehensive evaluation methodology and metric for TCPS. Towards this, we have developed an evaluation model and have argued that we can determine the quality of a TCPS by evaluating its critical control loop. Depending on the application, we propose evaluating either the  \textit{kinematic-haptic} or the \textit{kinematic-video} loop or both.  
Our evaluation methodology is an adaptation of the classical step response method. Here, we have replaced the human operator with a controller with known characteristics and have analyzed its response to slave side step changes.  We have described methods to simulate step-change in scenarios where there exist haptic feedback and where there is not. These methods extend the use case of our proposed evaluation methodology to a broader spectrum of TCPS implementations.  To quantify TCPS performance, we have developed a metric called \textit{Quality of Control (QoC)}. Through simulations and experiments on a TCPS testbed, we have demonstrated the practicality of the proposed evaluation methodology and its usefulness in quantifying cybersickness.

%\printbibliography
\bibliographystyle{IEEEtran}
\bibliography{myContents/references} 

% Generated by IEEEtran.bst, version: 1.14 (2015/08/26)
\begin{thebibliography}{10}
\providecommand{\url}[1]{#1}
\csname url@samestyle\endcsname
\providecommand{\newblock}{\relax}
\providecommand{\bibinfo}[2]{#2}
\providecommand{\BIBentrySTDinterwordspacing}{\spaceskip=0pt\relax}
\providecommand{\BIBentryALTinterwordstretchfactor}{4}
\providecommand{\BIBentryALTinterwordspacing}{\spaceskip=\fontdimen2\font plus
\BIBentryALTinterwordstretchfactor\fontdimen3\font minus
  \fontdimen4\font\relax}
\providecommand{\BIBforeignlanguage}[2]{{%
\expandafter\ifx\csname l@#1\endcsname\relax
\typeout{** WARNING: IEEEtran.bst: No hyphenation pattern has been}%
\typeout{** loaded for the language `#1'. Using the pattern for}%
\typeout{** the default language instead.}%
\else
\language=\csname l@#1\endcsname
\fi
#2}}
\providecommand{\BIBdecl}{\relax}
\BIBdecl

\bibitem{ref_polachanQoC}
K.~{Polachan}, P.~{T V}, C.~{Singh}, and D.~{Panchapakesan}, ``Quality of
  control assessment for tactile cyber-physical systems,'' in \emph{2019 16th
  Annual IEEE International Conference on Sensing, Communication, and
  Networking (SECON)}, June 2019, pp. 1--9.

\bibitem{6755599}
G.~P. Fettweis, ``{The Tactile Internet: Applications and Challenges},''
  \emph{IEEE Vehicular Technology Magazine}, vol.~9, no.~1, pp. 64--70, March
  2014.

\bibitem{ref_tactileInternetITU}
\BIBentryALTinterwordspacing
ITU. (2014) The tactile internet. [Online]. Available:
  \url{https://www.itu.int/en/ITU-T/techwatch/Pages/tactile-internet.aspx}
\BIBentrySTDinterwordspacing

\bibitem{ref_Arjun}
A.~{N}, A.~{S M}, K.~{Polachan}, P.~{T V}, and C.~{Singh}, ``An end to end
  tactile cyber physical system design,'' in \emph{2018 4th International
  Workshop on Emerging Ideas and Trends in the Engineering of Cyber-Physical
  Systems (EITEC)}, April 2018, pp. 9--16.

\bibitem{ref_LaViola2000}
\BIBentryALTinterwordspacing
J.~J. LaViola, Jr., ``{A Discussion of Cybersickness in Virtual
  Environments},'' \emph{SIGCHI Bull.}, vol.~32, no.~1, pp. 47--56, Jan. 2000.
  [Online]. Available: \url{http://doi.acm.org/10.1145/333329.333344}
\BIBentrySTDinterwordspacing

\bibitem{ref_aijaz2018tactile}
A.~Aijaz, Z.~Dawy, N.~Pappas, M.~Simsek, S.~Oteafy, and O.~Holland, ``{Toward a
  Tactile Internet Reference Architecture: Vision and Progress of the IEEE
  P1918.1 Standard},'' 2018.

\bibitem{ref_Steinbach}
K.~Antonakoglou, X.~Xu, E.~Steinbach, T.~Mahmoodi, and M.~Dohler, ``{Towards
  Haptic Communications over the 5G Tactile Internet},'' \emph{IEEE
  Communications Surveys Tutorials}, pp. 1--1, 2018.

\bibitem{ref_Kuipers2017}
D.~V.~D. Berg, R.~Glans, D.~D. Koning, F.~A. Kuipers, J.~Lugtenburg,
  K.~Polachan, P.~T. Venkata, C.~Singh, B.~Turkovic, and B.~V. Wijk,
  ``{Challenges in Haptic Communications Over the Tactile Internet},''
  \emph{IEEE Access}, vol.~5, pp. 23\,502--23\,518, 2017.

\bibitem{ref_Hamam}
A.~Hamam and A.~E. Saddik, ``{Evaluating the Quality of Experience of
  haptic-based applications through mathematical modeling},'' in \emph{2012
  IEEE International Workshop on Haptic Audio Visual Environments and Games
  (HAVE 2012) Proceedings}, Oct 2012, pp. 56--61.

\bibitem{ref_Kusunose}
Y.~Kusunose, Y.~Ishibashi, N.~Fukushima, and S.~Sugawara, ``{QoE assessment in
  networked air hockey game with haptic media},'' in \emph{2010 9th Annual
  Workshop on Network and Systems Support for Games}, Nov 2010, pp. 1--2.

\bibitem{ref_Jaafreh}
M.~A. Jaafreh, A.~Hamam, and A.~E. Saddik, ``{A framework to analyze fatigue
  for haptic-based tactile internet applications},'' in \emph{2017 IEEE
  International Symposium on Haptic, Audio and Visual Environments and Games
  (HAVE)}, Oct 2017, pp. 1--6.

\bibitem{ref_Tatematsu}
A.~Tatematsu, Y.~Ishibashi, N.~Fukushima, and S.~Sugawara, ``{QoE assessment in
  haptic media, sound and video transmission: Influences of network latency},''
  in \emph{2010 IEEE International Workshop Technical Committee on
  Communications Quality and Reliability (CQR 2010)}, June 2010, pp. 1--6.

\bibitem{ref_Sakr}
N.~Sakr, N.~D. Georganas, and J.~Zhao, ``{A Perceptual Quality Metric for
  Haptic Signals},'' in \emph{2007 IEEE International Workshop on Haptic, Audio
  and Visual Environments and Games}, Oct 2007, pp. 27--32.

\bibitem{ref_Correa}
C.~G. Corr\^{e}a, D.~M. Tokunaga, E.~Ranzini, F.~L.~S. Nunes, and R.~Tori,
  ``{Haptic Interaction Objective Evaluation in Needle Insertion Task
  Simulation},'' in \emph{Proceedings of the 31st Annual ACM Symposium on
  Applied Computing}, ser. SAC '16.\hskip 1em plus 0.5em minus 0.4em\relax New
  York, NY, USA: ACM, 2016, pp. 149--154.

\bibitem{ref_Chaudhari}
R.~Chaudhari, E.~Steinbach, and S.~Hirche, ``{Towards an objective quality
  evaluation framework for haptic data reduction},'' in \emph{2011 IEEE World
  Haptics Conference}, June 2011, pp. 539--544.

\bibitem{ref_Polachan}
K.~Polachan, T.~V. Prabhakar, C.~Singh, and F.~A. Kuipers, ``{Towards an Open
  Testbed for Tactile Cyber Physical Systems},'' in \emph{{11th International
  Conference on Communication Systems \& Networks, (COMSNETS)}}, 2019.

\bibitem{ref_Levitin}
D.~J. Levitin, K.~MacLean, M.~Mathews, L.~Chu, and E.~Jensen, ``{The perception
  of cross-modal simultaneity (or “the Greenwich Observatory Problem”
  revisited)},'' \emph{AIP Conference Proceedings}, vol. 517, no.~1, pp.
  323--329, 2000.

\bibitem{ref_Silva}
J.~M. Silva, M.~Orozco, J.~Cha, A.~E. Saddik, and E.~M. Petriu, ``{Human
  Perception of Haptic-to-video and Haptic-to-audio Skew in Multimedia
  Applications},'' \emph{ACM Trans. Multimedia Comput. Commun. Appl.}, vol.~9,
  no.~2, pp. 9:1--9:16, May 2013.

\bibitem{2452454}
A.~Aijaz, M.~Dohler, A.~H. Aghvami, V.~Friderikos, and M.~Frodigh, ``{Realizing
  the Tactile Internet: Haptic Communications over Next Generation 5G Cellular
  Networks},'' Accepted for publication in IEEE Wireless Communications, dec
  2015.

\bibitem{ref_Steinbach_HapticCodecs}
E.~{Steinbach}, M.~{Strese}, M.~{Eid}, X.~{Liu}, A.~{Bhardwaj}, Q.~{Liu},
  M.~{Al-Ja’afreh}, T.~{Mahmoodi}, R.~{Hassen}, A.~{El Saddik}, and
  O.~{Holland}, ``{Haptic Codecs for the Tactile Internet},''
  \emph{{Proceedings of the IEEE}}, vol. 107, no.~2, pp. 447--470, Feb 2019.

\bibitem{ref_Steinbach_hapticCommunications}
E.~Steinbach, S.~Hirche, M.~Ernst, F.~Brandi, R.~Chaudhari, J.~Kammerl, and
  I.~Vittorias, ``{Haptic communications},'' \emph{Proceedings of the IEEE},
  vol. 100, no.~4, pp. 937--956, 2012.

\bibitem{ref_SteinbachPPTHapticComm}
\BIBentryALTinterwordspacing
{Eckehard Steinbach}, ``{Haptic Communication for the Tactile Internet},''
  Tech. Rep., 2017. [Online]. Available:
  \url{http://ew2017.european-wireless.org/wp-uploads/2017/05/keynotes-Steinbach-EW2017.pdf}
\BIBentrySTDinterwordspacing

\bibitem{ref_ns3}
\BIBentryALTinterwordspacing
{ns-3}. (2018) {ns-3}. [Online]. Available: \url{https://www.nsnam.org}
\BIBentrySTDinterwordspacing

\bibitem{ref_mininet}
\BIBentryALTinterwordspacing
Mininet. (2019) {Mininet}. [Online]. Available: \url{http://mininet.org/}
\BIBentrySTDinterwordspacing

\bibitem{ref_Buttazzo2007}
G.~Buttazzo, M.~Velasco, and P.~Mart, ``{Quality-of-control management in
  overloaded real-time systems},'' \emph{IEEE Transactions on Computers},
  vol.~56, no.~2, pp. 253--266, 2007.

\bibitem{ref_Buttazzo2004}
G.~{Buttazzo}, M.~{Velasco}, P.~{Marti}, and G.~{Fohler}, ``Managing
  quality-of-control performance under overload conditions,''
  \emph{Proceedings. 16th Euromicro Conference on Real-Time Systems, 2004.
  ECRTS 2004.}, pp. 53--60, July 2004.

\bibitem{ref_Tian2011}
Y.~C. Tian and L.~Gui, ``{QoC elastic scheduling for real-time control
  systems},'' \emph{Real-Time Systems}, vol.~47, no.~6, pp. 534--561, 2011.

\bibitem{ref_Bini2008}
E.~Bini and A.~Cervin, ``{Delay-aware period assignment in control systems},''
  \emph{Proceedings - Real-Time Systems Symposium}, pp. 291--300, 2008.

\bibitem{ref_Aminifar2016}
A.~Aminifar, \emph{{Analysis, Design, and Optimization of Embedded Control
  Systems}}, 2016, no. 1746.

\bibitem{ref_Aminifar2018}
A.~Aminifar, P.~Eles, Z.~Peng, A.~Cervin, and K.~E. Arzen,
  ``{Control-Quality-Driven Design of Embedded Control Systems with Stability
  Guarantees},'' \emph{IEEE Design and Test}, vol.~35, no.~4, pp. 38--46, 2018.

\bibitem{ref_Cervin2002}
A.~Cervin, J.~Eker, B.~Bernhardsson, and K.~E. {\AA}rz{\'{e}}n,
  ``{Feedback-feedforward scheduling of control tasks},'' \emph{Real-Time
  Systems}, vol.~23, no. 1-2, pp. 25--53, 2002.

\bibitem{Fettweis:2014aa}
G.~Fettweis and S.~Alamouti, ``{5G: Personal mobile internet beyond what
  cellular did to telephony},'' \emph{IEEE Communications Magazine}, vol.~52,
  no.~2, pp. 140 -- 145, February 2014.

\bibitem{ref_vrep}
\BIBentryALTinterwordspacing
Coppeliarobotics. (2018) V-rep virtual robot experimental platform. [Online].
  Available: \url{http://www.coppeliarobotics.com/}
\BIBentrySTDinterwordspacing

\bibitem{ref_Gao2014}
Y.~Gao, S.~S. Vedula, C.~E. Reiley, N.~Ahmidi, B.~Varadarajan, H.~C. Lin,
  L.~Tao, L.~Zappella, B.~Bejar, D.~D. Yuh, C.~C.~G. Chen, R.~Vidal,
  S.~Khudanpur, and G.~D. Hager, ``Jhu isi gesture and skill assessment working
  set (jigsaws) a surgical activity dataset for human motion modeling,''
  \emph{Modeling and Monitoring of Computer Assisted Interventions (M2CAI) –
  MICCAI Workshop}, pp. 1--10, 2014.

\end{thebibliography}

\begin{comment}
\begin{IEEEbiography}{Kurian Polachan}
Biography text here.
\end{IEEEbiography}

\begin{IEEEbiography}{T V Prabhakar}
Biography text here.
\end{IEEEbiography}

\begin{IEEEbiography}{Chandramani Singh}
Biography text here.
\end{IEEEbiography}

\begin{IEEEbiography}{Deepak}
Biography text here.
\end{IEEEbiography}

\begin{IEEEbiography}{Joydee}
Biography text here.
\end{IEEEbiography}
\end{comment}
\end{document}